\documentclass[aps,prd,twocolumn,nofootinbib,longbibliography,10pt]{revtex4-2}
\usepackage{etoolbox}
\usepackage{dcolumn}
\usepackage{amsmath,amssymb,amsfonts,mathtools}
\usepackage{mathrsfs,bbold}
\usepackage{graphicx}
\usepackage[colorlinks=true,urlcolor=blue,citecolor=red,linkcolor=blue]{hyperref}
\usepackage{accents}
\newlength{\dhatheight}
\usepackage{natbib}
\usepackage{wrapfig}
\usepackage{flushend,BOONDOX-cal,BOONDOX-frak}

\makeatletter
\newsavebox{\@brx}
\newcommand{\llangle}[1][]{\savebox{\@brx}{\(\m@th{#1\langle}\)}%
  \mathopen{\copy\@brx\kern-0.5\wd\@brx\usebox{\@brx}}}
\newcommand{\rrangle}[1][]{\savebox{\@brx}{\(\m@th{#1\rangle}\)}%
  \mathclose{\copy\@brx\kern-0.5\wd\@brx\usebox{\@brx}}}
\makeatother
\newcommand{\doublehat}[1]{%
    \settoheight{\dhatheight}{\ensuremath{\hat{#1}}}%
    \addtolength{\dhatheight}{-0.35ex}%
    \hat{\vphantom{\rule{1pt}{\dhatheight}}%
    \smash{\hat{#1}}}}
\begin{document}
\title{Probing the quantum nature of gravity with a Bose-Einstein condensate}
\author{Soham Sen}
\email{sensohomhary@gmail.com}
\affiliation{Department of Astrophysics and High Energy Physics, S. N. Bose National Centre for Basic Sciences, JD Block, Sector-III, Salt Lake City, Kolkata-700 106, India}
\author{Sunandan Gangopadhyay}
\email{sunandan.gangopadhyay@gmail.com}
\affiliation{Department of Astrophysics and High Energy Physics, S. N. Bose National Centre for Basic Sciences, JD Block, Sector-III, Salt Lake City, Kolkata-700 106, India}
\begin{abstract}
\noindent The effect of noise induced by gravitons has been investigated using a Bose-Einstein condensate. The general complex scalar field theory with a quadratic self-interaction term has been considered in the presence of a gravitational wave. The gravitational wave perturbation is then considerd as a sum of discrete Fourier modes in the momentum space. Varying the action and making use of the principle of least action, one obtains two equations of motion corresponding to the gravitational perturbation and the time-dependent part of the pseudo-Goldstone boson. Coming to an operatorial representation and quantizing the phase space variables via appropriately introduced canonincal commutation relations between the canonically conjugate variables corresponding to the graviton and bosonic  part of the total system, one obtains a proper quantum gravity setup. Then we obtain the Bogoliubov coefficients from the solution of the time-dependent part of the pseudo-Goldstone boson and construct the covariance metric for the bosons initially being in a squeezed state. The entries of the covariance matrix now involves a stochastic contribution which results in an operatorial stochastic structure of the quantum Fisher information. Using the stochastic average of the Fisher information, we obtain a lower bound on the amplitude parameter of the gravitational wave. As the entire calculation is done at zero temperature, the bosonic system, by construction, will behave as a Bose-Einstein condensate. For a Bose-Einstein condensate with a single mode, we observe that the lower bound of the expectation value of the square of the uncertainty in the amplitude measurement does not become infinite when the total observational term approaches zero. It always has a finite value if the gravitons are initially in a squeezed state with high enough squeezing. In order to sum over all possible momentum modes, we next consider a noise term with a suitable Gaussian weight factor which decays over time. We then obtain the lower bound on the final expectation value of the square of the variance in the amplitude parameter. Because of the noise induced by the graviton, there is a minimum value of the measurement time below which it is impossible to detect any gravitational wave using a Bose-Einstein condensate. Finally, we consider interaction between the phonon modes of the Bose-Einstein condensate which results in a decoherence. We observe that the decoherence effect becomes significant for gravitons with minimal squeezing.
\end{abstract}
\maketitle
\section{Introduction}\label{S1}
\noindent The derivation of the Planck's radiation law by Satyendranath Bose in 1924 \cite{SNBose} led to the introduction of the Bose statistics. Albert Einstein in this time frame of 1924 and 1925 \cite{Einstein1,Einstein2,Einstein3} extended this idea to matter systems which led to the idea of a Bose gas governed by Bose statistics. Einstein also proposed the existence of a new state of matter which was later termed as Bose-Einstein condensate. The idea of a Bose-Einstein condensation is that if a bosonic system (even a boson gas) is cooled below a critical temperature all the bosons occupy the ground state energy level of the system. The matter waves start superposing when the de-Broglie wavelength is larger than the interatomic distance of the individual atoms and eventually at the moment of crossing the critical temperature, all the matter waves superpose to form a single wave function occupying the ground state of the bosonic system. This phenomena is termed as Bose-Einstein condensation. Experimentally, Bose-Einstein condensation was first detected in 1995 in a gas of Rubidium atoms \cite{1Nobel2001} and later in a gas of Sodium atoms in the same year \cite{2Nobel2001}. Since then people have tried to evolve the method for producing Bose-Einstein condensates and use it for various physical applications. Another important aspect of theoretical physics was the experimental detection of gravitational waves \cite{GWaveDetection, GWaveDetection2,GWaveDetection3} which has opened up a new era of theoretical physics involving the sculpturization of subatomic or lower physical length scales via the use of gravitational waves. Gravitational wave detection by using atom interferometry has been proposed quite a some time ago in \cite{AtomInterferometry,AtomInterferometry2}. Recently in \cite{PhononBEC}, a gravitational wave detector using a Bose-Einstein condensate has been proposed where a zero temperature quasi (1+1)-dimensional  Bose-Einstein condensate with fluctuating boundary conditions has been considered. An alternative calculation considering the interaction of a nonrelativistic Bose-Einstein condensate with a gravitational  wave has been done in \cite{PhononBEC2}. Later in \cite{PhononBEC3,PhononBEC4}, the idea propsed in \cite{PhononBEC} has been extended and made much more enhanced using a $(3+1)$-dimensional zero temperature Bose-Einstein condensation and a decaying gravitational wave template. The quantum Fisher information $H_\varepsilon$ was calculated by analyzing the fidelity between the individual squeezed phonon states. The quantum Fisher information carries the amount of information carried by the gravitational wave. Recently in \cite{HartleyKadingHowlFuentes}, a novel experimental setup has been proposed using BEC inteferometry to detect dark energy signatures in nature.

\noindent Recently, another aspect of high energy physics has emerged where the stochastic effect of the noise of gravitons from a linearized quantum gravity setup has been oberved \cite{QGravNoise,QGravLett,QGravD,
KannoSodaTokuda,KannoSodaTokuda2,AppleParikh,
OTMGraviton,OTMApple}. In \cite{QGravNoise,QGravLett,QGravD}, an interferometer detector has been modelled by means of a freely falling pair of particles in a slightly curved background. The perturbation over the Minkowski spacetime has then been decomposed into its discrete Fourier modes in $(1+1)$-dimension. Following a path integral approach the influence functional of the gravitons over the detector system was calculated and varying the action with respect to the detector degrees of freedom, the geodesic deviation equation was obtained, which had the structure of a Langevin-like stochastic differential equation. In these works, it has been shown that if a graviton is initially in a squeezed state then it may be possible to detect signatures from the detector-graviton interaction in future generation of gravitational wave detectors. Another set of analysis were done in $(3+1)$-spacetime dimensions and a canonical approach was followed. Similar stochastic Langevin-like equations were obtained \cite{KannoSodaTokuda} and an indirect detection of gravitons by means of decoherence was proposed in  \cite{KannoSodaTokuda2}. Similar but unique  stochastic effects has been observed in several other analyses \cite{AppleParikh,OTMGraviton,OTMApple}. The interaction between graviton and its possible detection scenarios as well as some important physical aspects have been quite thoroughly investigated in \cite{Novikov,SettiParikh,ChoHu}.

\noindent The primary motivation of this work is to unveil the effects of the noise induced by the gravitons on a homogeneous Bose-Einstein condensate in $(3+1)$-spacetime dimensions at zero temperature. To carry out the analysis, we need to start with the combined action comprising of the action describing the Bose-Einstein condensate in curved spacetime and the Einstein-Hilbert action. Here we have got rid of all the heavy fields in the theory as they will have very small contribution towards the overall dynamics of the theory. From \cite{PhononBEC3,PhononBEC4}, we already know that  a BEC is susceptible to gravitational wave when the resonance condition is matched. As the gravitational fluctuation in our analysis is now quantized, we expect to observe more subtle effects of the gravitons on the phonons. Such small effects can lead to a BEC state which will be incorporate such noise fluctuations. If one can now find a way to trace such signatures of fluctuations due to graviton-BEC interaction, it will suffice as an indirect detection of gravitons. In our analysis, we investigate the BEC-graviton interaction using quantum metrological techniques and we consider the quantum Fisher information to be the primary tool for indicating quantum gravity signatures. Due to such noise fluctuations, one needs to look at now the stochastic average of the quantum-gravity modified Fisher information. The square root of the stochastic average of the Fisher information will give the minimum value of the standard deviation in the amplitude parameter of the gravitational wave. We investigate the form of the quantum gravity modified Fisher information for squeezed graviton states which also will highlight the primary analysis of the paper. Later we have considered a scenario when the noise fluctuation is controlled by a Gaussian decay factor. Next, we investigate on whether the BEC will be a good candidate for extracting signatures of quantum gravity and compared it with required sensitivity data from space based gravitational wave observatories. We have finally investigated the effect of decoherence due to self interacting phonon modes in the quantum gravity modified Fisher information. People have also tried to invesigate observational effects of quantum gravity in interferometers \cite{ZurekVerlinde} and also have investigated modular fluctuations in shockwave geometries \cite{ZurekVerlinde2}. In \cite{ZurekVerlinde}, spacetime fluctuation in the arm of the interferometer detector is considered which is a direct consequence of the quantum nature of gravity. As gravitational wave interferometers are the best tools to detect very small fluctuations in the spacetime geometries the authors in \cite{ZurekVerlinde} have made use of the important infrared effects naturally arising from holography combined with the Planck scale fluctuations and proposed a indirect detection for quantum gravity signatures. This method can in principle provide an interesting testing ground of quantum gravity for BEC based gravitational wave detectors.  It would be interesting to see whether these proposals can be implemented with the BEC gravitational wave detectors.

\noindent Our paper is organized as follows. In section (\ref{S2}), we obtain the total action of the system. In section (\ref{S3}), we discuss the noise induced by the gravitons in the Bose-Einstein condensate. Later in section (\ref{S4}) we discuss the quantum metrology and obtain the quantum Fisher information for the system. We consider a different noise template in section (\ref{S5}). In section (\ref{S6}) we consider decoherence due to interacting phonon modes and finally in section \ref{S7}, we summarize our results.
\section{Action of the system} \label{S2}
\noindent In this section, we shall obtain the total action for the system in which a gravitational wave is interacting with a self interacting scalar field theory describing bosons. We work in the mostly positive signature for the metric. 
The background metric can be thought of as a small perturbation on the flat Minkowski background. The background metric is given by
\begin{equation}\label{1.1}
g_{\mu\nu}=\eta_{\mu\nu}+h_{\mu\nu}
\end{equation} 
where $\eta_{\mu\nu}=diag\{-1,1,1,1\}$. If we consider the speed of light to be unity, then the Einstein-Hilbert action can be written as
\begin{equation}\label{1.2}
S_{EH}=\frac{1}{16\pi G}\int d^4 x \sqrt{-g} R
\end{equation}
with $R$ being the Ricci scalar and $g=\text{det}(g_{\mu\nu})$. Up to quadratic order in the perturbation term in eq.(\ref{1.1}), we can recast the Einstein Hilbert action as follows
\begin{equation}\label{1.3}
\begin{split}
S_{EH}\simeq&\frac{1}{64\pi G}\int d^4x ~(h_{\mu\nu}\Box h^{\mu\nu}-h\Box h+2 h^{\mu\nu}\partial_\mu\partial_\nu h\\&-2h_{\mu\alpha}\partial_\kappa\partial^{\alpha}h^{\mu\kappa})~.\end{split}
\end{equation}
Now we shall make use of the gauge symmetry of the perturbation term given by
\begin{equation}\label{1.4}
h_{\mu\nu}=\bar{h}_{\mu\nu}+\partial_\mu\xi_\nu+\partial_\nu\xi_\mu~. 
\end{equation}
Using this, we now impose the transverse-traceless gauge conditions given by
\begin{align}\label{1.5}
\partial_{\kappa}\bar{h}^{\kappa\zeta}=0~,\bar{h}^{\kappa}_\kappa=0~,~k_\rho\bar{h}^{\rho\zeta}=0
\end{align}
with $k_\rho=\delta_\rho^{0}$ being a constant time-like vector. In the transverse-traceless gauge, the form of the Einstein-Hilbert action in eq.(\ref{1.3}) can be recast as
\begin{equation}\label{1.6}
S_{EH}=-\frac{1}{8\kappa^2}\int d^4 x~ \partial_{\kappa}\bar{h}_{ij}\partial^{\kappa}\bar{h}^{ij}
\end{equation}
where $\kappa=\sqrt{8\pi G}$. One can now make use of a Fourier mode decomposition of the fluctuation term $\bar{h}_{ij}$ inside a box of volume $V$ as
\begin{equation}\label{1.7}
\bar{h}_{ij}(t,\mathbf{x})=\frac{2\kappa}{\sqrt{V}}\sum_{\mathbf{k},s}h^s(t,\mathbf{k})e^{i\mathbf{k}\cdot\mathbf{x}}\epsilon^s_{ij}(\mathbf{k})~.
\end{equation}
It is imperative to know that $\bar{h}_{ij}(t,\mathbf{x})=\bar{h}_{ij}^{*}(t,\mathbf{x})$ as $\bar{h}_{ij}(t,\mathbf{x})$ is a real quantity. Now making use of the Fourier mode decomposition in eq.(\ref{1.7}) and the reality condition of the fluctuation term, we can recast the Einstein-Hilbert action in eq.(\ref{1.6}) as
\begin{equation}\label{1.8}
\begin{split}
S_{\text{EH}}=\frac{1}{2}\sum_{\textbf{k},s}\int dt\left(\bigr|\dot{h}^s(t,\mathbf{k})\bigr|^2-k^2\bigr|h^s(t,\mathbf{k})\bigr|^2\right)~.
\end{split}
\end{equation}
The Lagrangian density for a complex scalar bosonic field with a self interaction term (in natural units) can be written as
\begin{equation}\label{1.9}
\begin{split}
\mathcal{L}&=\nabla_\mu\phi^\dagger\nabla^\mu\phi
+m^2\phi^\dagger\phi+\lambda \left(\phi^\dagger\phi\right)^2\\
&=g^{\mu\nu}\partial_\mu\phi^\dagger(t,\mathbf{x})\partial_\nu\phi(t,\mathbf{x})
+m^2|\phi(t,\mathbf{x})|^2+\lambda \left|\phi(t,\mathbf{x})\right|^4
\end{split}
\end{equation}
where $m$ gives the mass of the bosonic field and $\lambda\left|\phi\right|^4
$ gives the self interaction term for the bosons. Now eq.(\ref{1.9}) effectively describes the  Lagrangian density of a Bose-Einstein condensate (BEC) as we are doing a zero-temperature field theory. Note that the Lagrangian density presented in eq.(\ref{1.9}) is a bit different from the one presented in \cite{PhononBEC3,PhononBEC4} as we are working explicitly in a mostly positive signature. 

\noindent We now consider a homogeneous BEC  and write $\phi$ as $\phi(t,\mathbf{x})=e^{i\chi(t,\mathbf{x})}\varphi(t,\mathbf{x})$, where $\chi(t,\mathbf{x})$ and $\varphi(t,\mathbf{x})$ are both real. Substituting this relation back in eq.(\ref{1.9}), we obtain the modified Lagrangian density as
\begin{equation}\label{1.10}
\begin{split}
\mathcal{L}=g^{\mu\nu}\partial_\mu\varphi\partial_\nu\varphi+\varphi^2g^{\mu\nu}\partial_\mu\chi\partial_\nu\chi
+m^2\varphi^2+\lambda\varphi^4~.
\end{split}
\end{equation}
Here, $\varphi(t,\mathbf{x})$ is the heavy field, hence we extremize the Lagrangian density in eq.(\ref{1.10}) with respect to $\varphi$ as
\begin{equation}\label{1.11}
\begin{split}
\frac{\partial\mathcal{L}}{\partial \varphi}=2\varphi g^{\mu\nu}\partial_\mu\chi\partial_\nu\chi+2m^2\varphi
+4\lambda\varphi^3&=0 \\
\implies 2\varphi\left(g^{\mu\nu}\partial_\mu\chi\partial_\nu\chi+m^2
+2\lambda\varphi^2\right)&=0~.
\end{split}
\end{equation}
As $\varphi$ is an arbitrary scalar field, it is possible to write down the extremization condition from eq.(\ref{1.11}) as
\begin{equation}\label{1.12}
\begin{split}
\varphi^2=-\frac{1}{2\lambda}\left(g^{\mu\nu}\partial_\mu\chi\partial_\nu\chi+m^2\right)~.
\end{split}
\end{equation}
Substituting the above relation back in the Lagrangian density in eq.(\ref{1.10}) we get,
\begin{equation}\label{1.13}
\begin{split}
\mathcal{L}&=g^{\mu\nu}\partial_\mu\varphi\partial_\nu\varphi+\varphi^2\left(g^{\mu\nu}\partial_\mu\chi\partial_\nu\chi+m^2\right)+\lambda(\varphi^2)^2\\
&=g^{\mu\nu}\partial_\mu\varphi\partial_\nu\varphi-\frac{1}{4\lambda}\left(g^{\mu\nu}\partial_\mu\chi\partial_\nu\chi+m^2\right)^2~.
\end{split}
\end{equation}
The primary focus of the analysis lies in the low frequency regime and as a result the heavy field $\varphi$ can be integrated out from the theory \cite{DTSon,PhononBEC3}. As a result, we can define a new Lagrangian density with an effective minus sign as
\begin{equation}\label{1.14}
\mathcal{L}_{\text{BEC}}=\frac{1}{4\lambda}\left(g_{\mu\nu}\partial^\mu\chi\partial^\nu\chi+m^2\right)^2~~.
\end{equation}
Corresponding to this new Lagrangian, one can write down the total action of the matter part of the system (which is the BEC coupled to the gravity) as
\begin{equation}\label{1.15}
S_{\text{BEC}}=\int d^4x\sqrt{-g}\mathcal{L}_{\text{BEC}}
\end{equation}
where $g=\text{det}[g_{\mu\nu}]$. If $\pi(t,\mathbf{x})\in\mathbb{R}$ denotes the BEC phonons then in terms of these pseudo-Goldstone bosons, we can express $\chi$ as
\begin{equation}\label{1.16}
\chi(t,\mathbf{x})=-\tilde{\sigma}t+\pi(t,\mathbf{x})=\tilde{\sigma}x_\mu\delta^{\mu}_{~0}+\pi(t,\mathbf{x})
\end{equation}
where $x^0=t$ and $x_0=g_{0\mu}x^\mu=-t$.

\noindent The background metric has the form given by
\begin{equation}\label{1.17}
g_{\mu\nu}=
\begin{pmatrix}
-1&0&0&0\\
0&1+h_+(t)&h_\times(t)&0\\
0&h_\times(t)&1-h_+(t)&0\\
0&0&0&1
\end{pmatrix}
\end{equation}
where $h_+$ and $h_\times$ denote the plus and cross polarizations of the gravitational wave, propagating in the $z$ direction. From eq.(\ref{1.17}), we obtain, $\sqrt{-g}=\sqrt{1-(h_+^2+h_\times^2)}\simeq 1+\mathcal{O}(h_{\mu\nu}h^{\mu\nu})$. 
Using the decomposition in eq.(\ref{1.16}) and using the expansion of $\sqrt{-g}$, we can recast the action in eq.(\ref{1.15}) as
\begin{equation}\label{1.18}
\begin{split}
S_{\text{BEC}}&\simeq\frac{1}{4\lambda}\int d^4x \left(g_{\mu\nu}(\tilde{\sigma}\delta^\mu_{~0}+\partial^\mu\pi)(\tilde{\sigma}\delta^\nu_{~0}+\partial^\nu\pi)+m^2\right)^2
\end{split}
\end{equation}
where we have kept $\sqrt{-g}$ upto the leading order term and neglected the $\mathcal{O}(h_{\mu\nu}h^{\mu\nu})$ contribution. One can neglect the higher order derivative terms \cite{PhononBEC3} while expanding eq.(\ref{1.18}). Hence, eq.(\ref{1.18}) can be recast in the following form
\begin{equation}\label{1.19}
\begin{split}
S_{\text{BEC}}=&\int \frac{d^4x}{2\lambda}\bigr[(3\tilde{\sigma}^2-m^2)\dot{\pi}^2-(\tilde{\sigma}^2-m^2)g_{ij}\partial^i\pi\partial^j\pi\bigr]\\&+\int \frac{d^4x}{4\lambda}\left[4\tilde{\sigma}(\tilde{\sigma}^2-m^2)\dot{\pi}+(\tilde{\sigma}^2-m^2)^2\right]\\
=&\int \frac{d^4x}{2\lambda}\bigr[(3\tilde{\sigma}^2-m^2)\dot{\pi}^2-(\tilde{\sigma}^2-m^2)g_{ij}\partial^i\pi\partial^j\pi\bigr]\\&+\int \frac{d^3x}{\lambda}\left[\tilde{\sigma}(\tilde{\sigma}^2-m^2)\pi\right]\\
\simeq& \int \frac{d^4x}{2\lambda}\bigr[(3\tilde{\sigma}^2-m^2)\dot{\pi}^2-(\tilde{\sigma}^2-m^2)g_{ij}\partial^i\pi\partial^j\pi\bigr]
\end{split}
\end{equation}
where in the second line of the above equation we have got rid of the non-dynamical contributions and in the final line we have made use of the fact that $\pi(t,\mathbf{x})$ vanishes at the boundary. We now make use of the ansatz for the pseudo-Goldstone boson 
\begin{equation}\label{pseudo_Goldstone}
\pi(t,\mathbf{x})=\sum_{\mathbf{k}_\beta}e^{i\mathbf{k}_\beta\cdot\mathbf{x}}\psi_{\mathbf{k}_\beta}(t)~.
\end{equation}
As $\pi(t,\mathbf{x})\in\mathbb{R}$, we know that $\pi(t,\mathbf{x})=\pi^*(t,\mathbf{x})$. This reality condition leads us to the relation $\sum_{\mathbf{k}_\beta}e^{i\mathbf{k}_\beta\cdot \mathbf{x}}\psi_{\mathbf{k}_\beta}(t)=\sum_{\mathbf{k}_\beta}e^{i\mathbf{k}_\beta\cdot \mathbf{x}}\psi^*_{-\mathbf{k}_\beta}(t).$ The above relation implies $\psi_{\mathbf{k}_\beta}(t)=\psi^*_{-\mathbf{k}_\beta}(t)$ $\forall ~\mathbf{k}_\beta$. Throughout the analysis, we have neglected the spatial dependence of the gravitational fluctuation (this assumption has also been adopted in the classical treatment of \cite{PhononBEC3}). Using the above condition and the Fourier mode decomposition of the gravitational fluctuation term from eq.(\ref{1.7}), eq.(\ref{1.19}) can be recast as
\begin{widetext}
\begin{equation}\label{1.20}
\begin{split}
S_{\text{BEC}}\simeq&\frac{1}{2\lambda}\int d^4x\left[(3\tilde{\sigma}^2-m^2)\dot{\pi}(t,\mathbf{x})\dot{\pi}^*(t,\mathbf{x})-(\tilde{\sigma}^2-m^2)\left(\eta_{ij}+\bar{h}_{ij}(t,0)\right)\partial^i\pi(t,\mathbf{x})\partial^j\pi^*(t,\mathbf{x})\right]\\
=&\frac{1}{2\lambda}\int d^4x\biggr[(3\tilde{\sigma}^2-m^2)\sum\limits_{\mathbf{k}_\beta,\mathbf{k}'_\beta}e^{i(\mathbf{k}_\beta-\mathbf{k}'_\beta)\cdot\mathbf{x}}\dot{\psi}_{\mathbf{k}_\beta}(t)\dot{\psi}^*_{\mathbf{k}'_\beta}(t)-(\tilde{\sigma}^2-m^2)\Bigr[\eta_{ij}+\frac{2\kappa}{\sqrt{V}}\sum\limits_{\mathbf{k},s}h_{\mathbf{k},s}(t)\epsilon^s_{ij}(\mathbf{k})\Bigr]\\&\times\sum\limits_{\mathbf{k}_\beta,\mathbf{k}'_\beta}(ik_{\beta}^i)(-i{k'}_\beta^j)e^{i(\mathbf{k}_\beta-\mathbf{k}'_\beta)\cdot\mathbf{x}}\psi_{\mathbf{k}_\beta}(t)\psi^*_{\mathbf{k}'_\beta}(t)\biggr]\\
=&\frac{1}{2\lambda}\int dt\biggr[(3\tilde{\sigma}^2-m^2)\sum\limits_{\mathbf{k}_\beta,\mathbf{k}'_\beta}\dot{\psi}_{\mathbf{k}_\beta}(t)\dot{\psi}^*_{\mathbf{k}'_\beta}(t)\int d^3x~e^{i(\mathbf{k}_\beta-\mathbf{k}'_\beta)\cdot\mathbf{x}}-(\tilde{\sigma}^2-m^2)\Bigr[\eta_{ij}+\frac{2\kappa}{\sqrt{V}}\sum\limits_{\mathbf{k},s}h_{\mathbf{k},s}(t)\epsilon^s_{ij}(\mathbf{k})\Bigr]\\&\times\sum\limits_{\mathbf{k}_\beta,\mathbf{k}'_\beta}(ik_{\beta}^i)(-i{k'}_\beta^j)\psi_{\mathbf{k}_\beta}(t)\psi^*_{\mathbf{k}'_\beta}(t)\int d^3x~e^{i(\mathbf{k}_\beta-\mathbf{k}'_\beta)\cdot\mathbf{x}}\biggr]\\
=&\frac{1}{2\lambda}\int dt\biggr[(3\tilde{\sigma}^2-m^2)\sum\limits_{\mathbf{k}_\beta,\mathbf{k}'_\beta}\dot{\psi}_{\mathbf{k}_\beta}(t)\dot{\psi}^*_{\mathbf{k}'_\beta}(t)V_\beta\delta_{\mathbf{k}_\beta,\mathbf{k}'_\beta}-(\tilde{\sigma}^2-m^2)\Bigr[\eta_{ij}+\frac{2\kappa}{\sqrt{V}}\sum\limits_{\mathbf{k},s}h_{\mathbf{k},s}(t)\epsilon^s_{ij}(\mathbf{k})\Bigr]\\&\times\sum\limits_{\mathbf{k}_\beta,\mathbf{k}'_\beta}k_{\beta}^i{k'}_\beta^j\psi_{\mathbf{k}_\beta}(t)\psi^*_{\mathbf{k}'_\beta}(t)V_\beta\delta_{\mathbf{k}_\beta,\mathbf{k}'_{\beta}}\biggr]\\
\implies S_{\text{BEC}}=&\frac{V_\beta}{2\lambda}\int dt\biggr[(3\tilde{\sigma}^2-m^2)\sum\limits_{\mathbf{k}_\beta}\bigr|\dot{\psi}_{\mathbf{k}_\beta}(t)\bigr|^2-(\tilde{\sigma}^2-m^2)\Bigr[\eta_{ij}+\frac{2\kappa}{\sqrt{V}}\sum\limits_{\mathbf{k},s}h_{\mathbf{k},s}(t)\epsilon^s_{ij}(\mathbf{k})\Bigr]\sum\limits_{\mathbf{k}_\beta}k_{\beta}^i{k}_\beta^j\bigr|\psi_{\mathbf{k}_\beta}(t)\bigr|^2\biggr]
\end{split}
\end{equation}
\end{widetext}
where in the penultimate line we have made use of the normalization condition for the pseudo-Goldstone bosons inside a box of volume $V_\beta$ as $\int d^3x~e^{i(\mathbf{k}_\beta-\mathbf{k}'_\beta)\cdot\mathbf{x}}=V_\beta\delta_{\mathbf{k}_\beta,\mathbf{k}'_\beta}$. Here, $\delta_{\mathbf{k}_\beta,\mathbf{k}'_\beta}$ is the abbreviated form of $\delta_{\mathbf{k}_\beta,\mathbf{k}'_\beta}\equiv\delta_{{k}^1_\beta,{k'}^{1}_\beta}\times\delta_{{k}^2_\beta,{k'}^{2}_\beta}\times\delta_{{k}^3_\beta,{k'}^{3}_\beta}$. Such a box normalization of the pseudo-Goldstone bosons is quite intuitive in a sense that experimentally a Bose-Einstein condensate is formed in a very confined region (\textit{e.g.} making use of harmonic trap potentials) which can resemble the shape of a small box.
From the dispersion relation of the BEC \cite{PhononBEC3}, we know that $\omega_\beta^2\simeq c_s^2k_\beta^2$, where $c_s^2=\frac{\tilde{\sigma}^2-m^2}{3\tilde{\sigma}^2-m^2}$ (denoting the square of the speed of sound) and $k_\beta\ll m$ with $c_s\ll 1$ (in natural units, $c=1$). We can further recast eq.(\ref{1.20}) as
\begin{widetext}
\begin{equation}\label{1.21}
\begin{split}
S_{\text{BEC}}=&\gamma_\beta\int dt\biggr[\sum\limits_{\mathbf{k}_\beta}\bigr|\dot{\psi}_{\mathbf{k}_\beta}(t)\bigr|^2-c_s^2\Bigr[\eta_{ij}+\frac{2\kappa}{\sqrt{V}}\sum\limits_{\mathbf{k},s}h_{\mathbf{k},s}(t)\epsilon^s_{ij}(\mathbf{k})\Bigr]\sum\limits_{\mathbf{k}_\beta}k_\beta^ik_\beta^j\bigr|\psi_{\mathbf{k}_\beta}(t)\bigr|^2\biggr]
\end{split}
\end{equation}
where $\gamma_\beta$ has the dimension of length in natural units and $\gamma_\beta\equiv\frac{V_\beta}{2\lambda}(3\tilde{\sigma}^2-m^2)$. Combining the Einstein-Hilbert action from eq.(\ref{1.8}) with that of the action for the BEC from eq.(\ref{1.21}), we obtain the total action of the system as
\begin{equation}\label{1.22}
\begin{split}
S=&S_{\text{EH}}+S_{\text{BEC}}\\
=&\frac{1}{2}\sum_{\textbf{k},s}\int dt\left(\bigr|\dot{h}^s(t,\mathbf{k})\bigr|^2-k^2\bigr|h^s(t,\mathbf{k})\bigr|^2\right)+\gamma_\beta\int dt\sum\limits_{\mathbf{k}_\beta}\biggr[\bigr|\dot{\psi}_{\mathbf{k}_\beta}(t)\bigr|^2-c_s^2\Bigr[\eta_{ij}+\frac{2\kappa}{\sqrt{V}}\sum\limits_{\mathbf{k},s}h_{\mathbf{k},s}(t)\epsilon^s_{ij}(\mathbf{k})\Bigr]k_\beta^ik_\beta^j\bigr|\psi_{\mathbf{k}_\beta}(t)\bigr|^2\biggr]~.
\end{split}
\end{equation}
\end{widetext}
We start by varying the action given in eq.(\ref{1.22}) in terms of the complex conjugate of the  time dependent part of the pseudo-Goldstone boson corresponding to individual momentum modes and the complex conjugate of the individual Fourier mode of the graviton. Using the principle of least action ($\frac{\delta S}{\delta \psi^*_{\mathbf{k}_\beta}}=0$), we obtain a dynamical equation or the equation of motion corresponding to the time dependent part of the pseudo-Goldstone boson as
\begin{equation}\label{1.23}
\begin{split}
\ddot{\psi}_{\mathbf{k}_\beta}(t)+c_s^2\Bigr[\eta_{ij}+\frac{2\kappa}{\sqrt{V}}\sum\limits_{\mathbf{k},s}h_{\mathbf{k},s}(t)\epsilon^s_{ij}(\mathbf{k})\Bigr]{k}^i_\beta {k}^j_\beta\psi_{\mathbf{k}_\beta}(t)&=0~.
\end{split}
\end{equation}
Similarly, by using the principle of least action after extremixing the action with respect to the variable $h^*_{\mathbf{k},s}$, we get
\begin{equation}\label{1.24}
\begin{split}
\ddot{h}_{\mathbf{k},s}(t)+k^2h_{\mathbf{k},s}(t)=-\frac{4\gamma_\beta\kappa c_s^2}{\sqrt{V}}\epsilon^{s*}_{ij}(\mathbf{k})\sum\limits_{\mathbf{k}_\beta} k_\beta^ik_\beta^j\left\lvert\psi_{\mathbf{k}_\beta}(t)\right\rvert^2
\end{split}
\end{equation}
where we have made use of the reality condition of $\bar{h}_{ij}(t,0)$. With the two equations of motion in hand, we can now move towards writing down a quantum mechanical model of the gravitational wave-BEC system.
\section{Graviton induced noise in the BEC}\label{S3}
\noindent In this section, our primary aim is to quantize the theory. The simplest way of quantizing the gravitational part is to raise the Fourier modes of the spacetime fluctuation $\bar{h}_{ij}$ to operator status and impose a suitable canonical commutation relation among $\hat{h}_{\mathbf{k},s}(t)$ and its canonically conjugate variable in the phase space. For the bosonic part, it can be a bit tricky. One can just raise the time dependent part of the pseudo-Goldstone bosons to operator status and impose suitable commutation relation among the two canonically conjugate variables in the phase space. The second way is to quantize it in the momentum space and raise the momentum variables to operator status with the suitable use of canonical commutation relations. In this first paper of our set of two works, we shall make use of the first procedure, where we only quantize $\psi_{\mathbf{k}_\beta}(t)$ in its entirety and impose commutation relation between $\hat{\psi}_{\mathbf{k}_\beta}(t)$ and its canonically conjugate variable. The independent Fourier mode of the spacetime fluctuation can be decomposed into two parts. One is the classical contribution which is obtained by taking the expectation value of the mode operator and a quantum fluctuation term. It is now possible to write down the fluctuation term of the mode operator corresponding to a momentum value $\mathbf{k}$ in the interaction picture as \cite{KannoSodaTokuda}
\begin{equation}\label{1.25}
\delta \hat{h}^I_{\mathbf{k},s}(t)=\hat{h}_{\mathbf{k},s}^I(t)-h_{\text{cl}}^s(\mathbf{k},t)
\end{equation}
with  the classical component given as $h_{\text{cl}}^s(\mathbf{k},t)=\bigr\langle\hat{h}_{\mathbf{k},s}^I(t)\bigr\rangle$ where the expectation is taken with respect to the initial state of the graviton. In principle the classical contribution has non-vanishing contribution if the graviton is initially in coherent, squeezed vacuum, thermal, and similar other combinatorial states. Here, $\delta \hat{h}^I_{\mathbf{k},s}(t)$ is the gravitational quantum fluctuation \cite{KannoSodaTokuda}. The quantum field in the interaction picture in terms of the creation and annihilation operators can be represented as
\begin{equation}\label{1.26}
\hat{h}_{\mathbf{k},s}^I(t)=u_k(t)\hat{a}_s(\mathbf{k})+u_k^*(t)\hat{a}^\dagger_s(-\mathbf{k})
\end{equation} 
where $k=|\mathbf{k}|$ with the mode function, $u_k(t)$, satisfying the following normalization condition
\begin{equation}\label{1.27}
-iu_k(t)\overset{\leftrightarrow}{\partial}_tu_k^*(t)=-i\left(u_k(t)\dot{u}_k^*(t)-\dot{u}_k(t)u_k^*(t)\right)=1~.
\end{equation}
In case of Minkowski vacuum, the creation and annihilation operators in eq.(\ref{1.26}), satisfy the following commutation relation
\begin{equation}\label{1.28}
\begin{split}
[\hat{a}_s(\mathbf{k}),\hat{a}^\dagger_{s'}(\mathbf{k}')]&=\delta_{s,s'}\delta_{\mathbf{k},\mathbf{k}'},\\
[\hat{a}_s(\mathbf{k}),\hat{a}_{s'}(\mathbf{k}')]&=[\hat{a}^\dagger_s(\mathbf{k}),\hat{a}^\dagger_{s'}(\mathbf{k}')]=0~.
\end{split}
\end{equation}
Raising $h_{\mathbf{k},s}(t)$ and $\hat{\psi}_{\mathbf{k}_\beta}(t)$ to operator status, we can recast eq.(\ref{1.24}) in a quantum mechanical representation as
\begin{equation}\label{1.29}
\begin{split}
\ddot{\hat{h}}_{\mathbf{k},s}(t)+k^2\hat{h}_{\mathbf{k},s}(t)=-\frac{4\gamma_\beta\kappa c_s^2}{\sqrt{V}}\epsilon^{s*}_{ij}(\mathbf{k})\sum\limits_{\mathbf{k}_\beta} k_\beta^ik_\beta^j\left\lvert\hat{\psi}_{\mathbf{k}_\beta}(t)\right\rvert^2
\end{split}
\end{equation} 
where $\left\lvert\hat{\psi}_{\mathbf{k}_\beta}(t)\right\rvert^2=\hat{\psi}_{\mathbf{k}_\beta}^\dagger(t) \hat{\psi}_{\mathbf{k}_\beta}(t)$. 
\begin{widetext}
Making use of the Green's function technique, it is possible to write down the solution of the above equation as
\begin{equation}\label{1.30}
\begin{split}
\hat{h}_{\mathbf{k},s}(t)=&\hat{h}^I_{\mathbf{k},s}(t)-\frac{4\gamma_\beta\kappa c_s^2}{\sqrt{V}}\epsilon^{s*}_{ij}(\mathbf{k})\int_0^t dt'\frac{\sin(k(t-t'))}{k}\sum\limits_{\mathbf{k}'_\beta}{{k}'_\beta}^i {{k}'_\beta}^j\bigr|\hat{\psi}_{\mathbf{k}'_\beta}(t')\bigr|^2\\
=&h_{\text{cl}}^s(\mathbf{k},t)+\delta \hat{h}^I_{\mathbf{k},s}(t)-\frac{4\gamma_\beta\kappa c_s^2}{\sqrt{V}}\epsilon^{s*}_{ij}(\mathbf{k})\sum\limits_{\mathbf{k}'_\beta}{k'_\beta}^i{k'_\beta}^j\int_0^t dt'\frac{\sin(k(t-t'))}{k}\bigr|\hat{\psi}_{\mathbf{k}'_\beta}(t')\bigr|^2~.
\end{split}
\end{equation}
\end{widetext}
We can also write down the quantum mechanical version of  eq.(\ref{1.23}) as 
\begin{equation}\label{1.31}
\begin{split}
\ddot{\hat{\psi}}_{\mathbf{k}_\beta}(t)+c_s^2\Bigr[\eta_{ij}+\frac{2\kappa}{\sqrt{V}}\sum\limits_{\mathbf{k},s}\hat{h}_{\mathbf{k},s}(t)\epsilon^s_{ij}(\mathbf{k})\Bigr]{k}^i_\beta {k}^j_\beta\hat{\psi}_{\mathbf{k}_\beta}(t)&=0
\end{split}
\end{equation} 
where in the last line of the above equation, we have made use of eq.(\ref{1.25}). One can now regulate the mode summations, corresponding to the gravitational wave part, via the use of an ultraviolet (UV) cut-off. Substituting the form of $\hat{h}_{\mathbf{k},s}(t)$ from eq.(\ref{1.30}), we can write down eq.(\ref{1.31}) as
\begin{widetext}
\begin{equation}\label{1.32}
\begin{split}
&\ddot{\hat{\psi}}_{\mathbf{k}_\beta}(t)+c_s^2\eta_{ij}k_\beta^i k_\beta^j\hat{\psi}_{\mathbf{k}_\beta}(t)+c_s^2\biggr(\frac{2\kappa }{\sqrt{V}}\sum\limits_{s}\smashoperator{\sum\limits_{\substack{{\mathbf{k}}\\{|\mathbf{k}|\leq\Omega_m}}}}h_{\text{cl}}^s(\mathbf{k},t)\epsilon^s_{ij}(\mathbf{k})\biggr)k_\beta^ik_\beta^j\hat{\psi}_{\mathbf{k}_\beta}(t)+c_s^2\biggr(\frac{2\kappa }{\sqrt{V}}\sum\limits_{s}\smashoperator{\sum\limits_{\substack{{\mathbf{k}}\\{|\mathbf{k}|\leq\Omega_m}}}}\delta \hat{h}^I_{\mathbf{k},s}(t)\epsilon^s_{ij}(\mathbf{k})\biggr)k_\beta^ik_\beta^j\hat{\psi}_{\mathbf{k}_\beta}(t)\\&-\frac{8\gamma_\beta \kappa^2c_s^4}{V}\sum\limits_{s}\smashoperator{\sum\limits_{\substack{{\mathbf{k}}\\{|\mathbf{k}|\leq\Omega_m}}}}\epsilon^{s*}_{ij}(\mathbf{k})\epsilon^s_{lm}(\mathbf{k})\biggr(\int_0^t dt'\frac{\sin(k(t-t'))}{k}\sum\limits_{\mathbf{k}'_\beta}{k'_\beta}^i{k'_\beta}^j\bigr|\hat{\psi}_{\mathbf{k}'_\beta}(t')\bigr|^2\biggr)k_\beta^lk_\beta^m\hat{\psi}_{\mathbf{k}_\beta}(t)=0~.
\end{split}
\end{equation}
\end{widetext}
We will henceforth use the UV-regulated mode summations for the gravitational wave part throughout this work. From an experimental scenario this is quite logical as an usual gravitational wave detector can detect frequencies upto a maximum value. 

\noindent One can now define two new quantities  as
\begin{align}
h_{ij}^{\text{cl}}(t,\mathbf{x})\equiv&\frac{2\kappa}{\sqrt{V}}\sum\limits_{s}\smashoperator{\sum\limits_{\substack{{\mathbf{k}}\\{|\mathbf{k}|\leq\Omega_m}}}}h_{\text{cl}}^s(\mathbf{k},t)e^{i\mathbf{k}\cdot \mathbf{x}}\epsilon^s_{ij}(\mathbf{k})\label{1.33}\\
\delta\hat{N}_{ij}(t)\equiv&\frac{2\kappa }{\sqrt{V}}\sum\limits_{s}\smashoperator{\sum\limits_{\substack{{\mathbf{k}}\\{|\mathbf{k}|\leq\Omega_m}}}}\delta \hat{h}^I_{\mathbf{k},s}(t)\epsilon^s_{ij}(\mathbf{k})~.\label{1.34}
\end{align}
One can now introduce the projection tensors to proceed further in this analysis as follows
\begin{equation}\label{1.35}
\begin{split}
\mathcal{P}_{ij}=\delta_{ij}-\frac{k_ik_j}{k^2}
\end{split}
\end{equation}
and making use of them, one can write down the following relation involving the sum over all polarizations of the product of two polarization tensors
\begin{equation}\label{1.36}
\begin{split}
\sum\limits_{s}\epsilon^{s*}_{ij}(\mathbf{k})\epsilon^s_{lm}(\mathbf{k})=\frac{1}{2}\left[\mathcal{P}_{il}\mathcal{P}_{jm}+\mathcal{P}_{im}\mathcal{P}_{jl}-\mathcal{P}_{ij}\mathcal{P}_{lm}\right]~.
\end{split}
\end{equation}
Using eq.(s)(\ref{1.33},\ref{1.34}) and eq.(\ref{1.36}), we can recast eq.(\ref{1.32}) as 
\begin{equation}\label{1.37}
\begin{split}
&\ddot{\hat{\psi}}_{\mathbf{k}_\beta}(t)+c_s^2\left(\eta_{ij}+h_{ij}^{\text{cl}}(t,0)+\delta\hat{N}_{ij}(t)\right)k_\beta^i k_\beta^j\hat{\psi}_{\mathbf{k}_\beta}(t)\\&-\frac{\xi_\beta}{V}\smashoperator{\sum\limits_{\substack{{\mathbf{k}}\\{|\mathbf{k}|\leq\Omega_m}}}}\left(\mathcal{P}_{il}\mathcal{P}_{jm}+\mathcal{P}_{im}\mathcal{P}_{jl}-\mathcal{P}_{ij}\mathcal{P}_{lm}\right)\sum\limits_{\mathbf{k}'_\beta}{k'_\beta}^i{k'_\beta}^j\\&\times\left(\int_0^t dt'\frac{\sin(k(t-t'))}{k}\bigr|\hat{\psi}_{\mathbf{k}'_\beta}(t')\bigr|^2\right)k_\beta^lk_\beta^m\hat{\psi}_{\mathbf{k}_\beta}(t)=0
\end{split}
\end{equation}
where $\xi_\beta=4\gamma_\beta \kappa^2c_s^4$, and $\frac{\xi_\beta}{V}$ is a dimensionless number. The summation over the graviton modes can be converted in a continuous mode integral as $\frac{1}{V}\smashoperator{\sum_{{\mathbf{k}};{|\mathbf{k}|\leq\Omega_m}}}\rightarrow\frac{1}{(2\pi)^3}\int^{\Omega_m}d^3k$. In this integral $d^3k$ can be converted to corresponding spherical coordinates in Fourier space as $k^2 dk \sin\theta d\theta d\phi=k^2 dk d\Omega$. The angular integrals are given by \cite{KannoSodaTokuda}
\begin{equation}\label{1.38}
\begin{split}
&\int d\Omega=4\pi,~\int d\Omega ~k^ik^j=\frac{4\pi}{3}\delta_{ij},\\
&\int d\Omega~k^ik^jk^lk^m =\frac{4\pi}{15}\left(\delta^{ij}\delta^{lm}+\delta^{il}\delta^{jm}+\delta^{im}\delta^{jl}\right).
\end{split}
\end{equation}
Using the angular integrals and making use of the discrete to continuous mode conversion rule, we obtain the following result for the summation terms involving the projection tensors, $\sum_\mathbf{k}\left(\mathcal{P}_{il}\mathcal{P}_{jm}+\mathcal{P}_{im}\mathcal{P}_{jl}-\mathcal{P}_{ij}\mathcal{P}_{lm}\right)$, to be
\begin{widetext}
\begin{equation}\label{1.39}
\begin{split}
&\frac{V}{(2\pi)^3}\int^{\Omega_m}_0dk  k^2 \int d\Omega\left[\mathcal{P}_{il}\mathcal{P}_{jm}+\mathcal{P}_{im}\mathcal{P}_{jl}-\mathcal{P}_{ij}\mathcal{P}_{lm}\right]=\frac{V}{5\pi^2}\left(\int_0^{\Omega_m}dk k^2\right)\left(\delta_{il}\delta_{jm}+\delta_{im}\delta_{jl}-\frac{2}{3}\delta_{ij}\delta_{lm}\right)~.
\end{split}
\end{equation}
Making use of eq.(\ref{1.39}) and performing the $k$ integral, one can recast eq.(\ref{1.37}) as
\begin{equation}\label{1.40}
\begin{split}
&\ddot{\hat{\psi}}_{\mathbf{k}_\beta}(t)+c_s^2\left(\eta_{ij}+h_{ij}^{\text{cl}}(t,0)+\delta\hat{N}_{ij}(t)\right)k_\beta^i k_\beta^j\hat{\psi}_{\mathbf{k}_\beta}(t)-\frac{\xi_\beta}{5\pi^2}\left(\delta_{il}\delta_{jm}+\delta_{im}\delta_{jl}-\frac{2}{3}\delta_{ij}\delta_{lm}\right)\sum\limits_{\mathbf{k}'_\beta}{k'_\beta}^i{k'_\beta}^j\\
&\times\int_0^t dt'\left(\frac{\sin(\Omega_m(t-t'))}{(t-t')^2}-\frac{\Omega_m\cos(\Omega_m(t-t'))}{(t-t')}\right)\bigr|\hat{\psi}_{\mathbf{k}'_\beta}(t')\bigr|^2k_\beta^lk_\beta^m\hat{\psi}_{\mathbf{k}_\beta}(t)=0~.
\end{split}
\end{equation}
Absorbing the Kronecker-deltas, we can resimplify the above equation as
\begin{equation}\label{1.41}
\begin{split}
&\ddot{\hat{\psi}}_{\mathbf{k}_\beta}(t)+c_s^2\left(\eta_{ij}+h_{ij}^{\text{cl}}(t,0)+\delta\hat{N}_{ij}(t)\right)k_\beta^i k_\beta^j\hat{\psi}_{\mathbf{k}_\beta}(t)-\frac{2\xi_\beta}{5\pi^2}\int_0^tdt'\left(\frac{\sin(\Omega_m(t-t'))}{(t-t')^2}-\frac{\Omega_m\cos(\Omega_m(t-t'))}{(t-t')}\right)\\&\times\sum_{\mathbf{k}'_\beta}\left((\mathbf{k}_\beta\cdot\mathbf{k}'_\beta)^2-\frac{1}{3}k_\beta^2{k'}^2_\beta\right)\bigr|\hat{\psi}_{\mathbf{k}'_\beta}(t')\bigr|^2\hat{\psi}_{\mathbf{k}_\beta}(t)=0~.
\end{split}
\end{equation}
\end{widetext}
Our next aim is to obtain a solution for $\hat{\psi}_{\mathbf{k}_\beta}(t)$ via solving the above equation. The time-dependent part of the pseudo-Goldstone boson can be divided into three parts, $\hat{\psi}_{\mathbf{k}_\beta}(t)=\psi^{(0)}_{\mathbf{k}_\beta}(t)+\psi^{h_{\text{cl}}}_{\mathbf{k}_\beta}(t)+\hat{\psi}^{(1)}_{\mathbf{k}_\beta}(t)$. Here, $\psi_{\mathbf{k}_\beta}^{(0)}(t)$ denotes the unperturbed classical part of the solution and $\psi_{\mathbf{k}_\beta}^{h_{\text{cl}}}(t)$ denotes the first order solution corresponding to the classical gravitational perturbation. The final part $\hat{\psi}^{(1)}_{\mathbf{k}_\beta}(t)$ encodes the solution corresponding to the quantum fluctuations of the gravitons. It is important to note that the decomposition of the solution is done in a way such that the operatorial contribution can be separated from the classical part. 

\noindent The zeroth order classical equation from eq.(\ref{1.41}) reads
\begin{equation}\label{1.42}
\begin{split}
\ddot{\psi}^{(0)}_{\mathbf{k}_\beta}(t)+c_s^2k_\beta^2\psi^{(0)}_{\mathbf{k}_\beta}(t)=0
\end{split}
\end{equation}
which has a solution of the form
\begin{equation}\label{1.43}
\begin{split}
\psi^{(0)}_{\mathbf{k}_\beta}(t)&=\mathcal{a} e^{-i c_sk_\beta t}+\mathcal{b} e^{i c_sk_\beta t}=\mathcal{a} e^{-i \omega_\beta t}+\mathcal{b} e^{i \omega_\beta t}~.
\end{split}
\end{equation}
It is now quite intuitive to get rid of the negative energy modes and set $\mathcal{b}=0$, and as a result of the normalization condition, we get $\mathcal{a}=1$. The first order classical equation of motion from eq.(\ref{1.41}) can be written as
\begin{equation}\label{1.44}
\begin{split}
\ddot{\psi}^{h_{\text{cl}}}_{\mathbf{k}_\beta}(t)+c_s^2k_\beta^2{\psi}^{h_{\text{cl}}}_{\mathbf{k}_\beta}(t)=-c_s^2h^{\text{cl}}_{ij}(t,0)k_\beta^ik_\beta^j\psi^{(0)}_{\mathbf{k}_\beta}(t)+\mathcal{f}_\beta(t)
\end{split}
\end{equation}
where $\mathcal{f}_\beta(t)$ is given by
\begin{equation}\label{1.45}
\begin{split}
\mathcal{f}_\beta(t)&=\frac{2\xi_\beta}{5\pi^2}\sum_{\mathbf{k}'_\beta}\left[(\mathbf{k}_\beta\cdot\mathbf{k}'_\beta)^2-\frac{1}{3}k_\beta^2{k'}^2_\beta\right]\psi^{(0)}_{\mathbf{k}_\beta}(t)\int_0^tdt'\\\times&\biggr[\frac{\sin(\Omega_m(t-t'))}{(t-t')^2}-\frac{\Omega_m\cos(\Omega_m(t-t'))}{(t-t')}\biggr]\bigr|{\psi}^{(0)}_{\mathbf{k}'_\beta}(t')\bigr|^2\\
&=\frac{2\xi_\beta}{5\pi^2}\left[\Omega_m-\frac{\sin(\Omega_m t)}{t}\right]\sum_{\mathbf{k}'_\beta}\Bigr[(\mathbf{k}_\beta\cdot\mathbf{k}'_\beta)^2\\&-\frac{1}{3}k_\beta^2{k'}^2_\beta\Bigr]e^{-i\omega_\beta t}~.
\end{split}
\end{equation}
To proceed further and in order to make the analysis simpler, we restrict ourselves to plus polarization of the gravitational wave only. As a result, we already know that $\epsilon^\times_{ij}(\mathbf{k})=0$ ($\forall i,j=\{1,2,3\}$), $\epsilon^+_{11}(\mathbf{k})=-\epsilon^+_{22}(\mathbf{k})$, $\epsilon^+_{33}(\mathbf{k})=0$, and $\epsilon^+_{ij}(\mathbf{k})=0$ $\forall i\neq j$. Making use of eq.(\ref{1.33}), we can therefore write $h^{\text{cl}}_{11}(t,0)=-h^{\text{cl}}_{22}(t,0)=h^{\text{cl}}(t,0)$. Hence, eq.(\ref{1.44}) can be recast as
\begin{equation}\label{1.46}
\begin{split}
\ddot{\psi}^{h_{\text{cl}}}_{\mathbf{k}_\beta}(t)+\omega_\beta^2\psi^{h_{\text{cl}}}_{\mathbf{k}_\beta}(t)&=-\omega_\beta^2\mathcal{k}_0^2h^{\text{cl}}(t,0)e^{-i\omega_\beta t}+\mathcal{f}_\beta(t)
\end{split}
\end{equation}
where $\mathcal{k}_0^2=\frac{{k_\beta}_x^2-{k_\beta}_y^2}{k_\beta^2}$. Corresponding to the Green's function equation $(\frac{d^2}{dt^2}+\omega_\beta^2)\mathcal{G}(t-t')=\delta(t-t)'$, we obtain the analytical form of the Green's function  to be
\begin{equation}\label{1.47}
\mathcal{G}(t-t')=\frac{1}{\omega_\beta}\sin(\omega_\beta(t-t'))\Theta(t-t')
\end{equation}
with $\Theta(t-t')$ denoting the Heaviside theta function. The solution of eq.(\ref{1.44}) can then be obtained as
\begin{equation}\label{1.48}
\begin{split}
&\psi^{h_{\text{cl}}}_{\mathbf{k}_\beta}(t)=\mathcal{a}_he^{-i\omega_\beta t}+\mathcal{b}_he^{i\omega_\beta t}-\mathcal{k}_0^2\int_{-\infty}^tdt'\omega_\beta h^{\text{cl}}(t',0)\\&\times e^{-i\omega_\beta t'}\sin(\omega_\beta[t-t'])+\int_{-\infty}^tdt'\frac{\sin(\omega_\beta[t-t'])}{\omega_\beta}\mathcal{f}_\beta(t')~.
\end{split}
\end{equation}
The simplest choice for the undetermined constants is to set $\mathcal{a}_h=\mathcal{b}_h=0$. As before, we can also get $\delta\hat{N}_{11}(t)=-\delta\hat{N}_{22}(t)=\delta\hat{N}(t)$ and other components of the $\delta\hat{N}_{ij}(t)$ tensor is zero. The final dynamical  equation, involving the operators only (the quantum-gravitational time evolution equation), reads
\begin{equation}\label{1.49}
\begin{split}
\ddot{\hat{\psi}}^{(1)}_{\mathbf{k}_\beta}(t)+\omega_\beta^2\hat{\psi}^{(1)}_{\mathbf{k}_\beta}(t)\simeq-\omega_\beta^2\mathcal{k}_0^2\delta\hat{N}(t).
\end{split}
\end{equation}
As $\hat{\psi}^{(1)}_{\mathbf{k}_\beta}(t)$ is an operator corresponding to quantum gravity consideration, the operatorial contribution from $\mathcal{f}_\beta(t)$ becomes way smaller compared to the other terms in eq.(\ref{1.49}). Again setting the random constants to zero and using the Green's function technique, we arrive at the solution of eq.(\ref{1.49}) as follows
\begin{equation}\label{1.50}
\begin{split}
\hat{\psi}^{(1)}_{\mathbf{k}_\beta}(t)=-\mathcal{k}_0^2\int_{-\infty}^tdt'\omega_\beta\sin(\omega_\beta(t-t'))e
^{-i\omega_\beta t'}\delta\hat{N}(t')~.\end{split}
\end{equation}
As the quantum fluctuations purely arises because of the interaction of the BEC with the gravitons, it is safe to assume $\delta\hat{N}(t)=0$, $\forall t<0$. As a result, eq.(\ref{1.50}) reduces to an integral whose limits are from $0$ to $t$. Combining eq.(s)(\ref{1.43},\ref{1.46}) and eq.(\ref{1.50}) and using the specific values for the constant, we obtain the complete solution for the time dependent part of the pseudo-Goldstone boson as
\begin{widetext}
\begin{equation}\label{1.51}
\hat{\psi}_{\mathbf{k}_\beta}(t)=e^{-i\omega_\beta t}-\omega_\beta\mathcal{k}_0^2\int^t_{-\infty}dt'e^{-i\omega_\beta t'}\sin(\omega_\beta(t-t'))\left(h^{\text{cl}}(t',0)+\delta\hat{N}(t')\right)+\frac{1}{\omega_\beta}\int^t_{-\infty}dt'\sin(\omega_\beta(t-t'))\mathcal{f}_\beta(t')~.
\end{equation}
This is one of the pivotal results in our paper and it signifies the fact that the time dependent part of the pseudo-Goldstone boson now explicitly depends upon a quantum fluctuation parameter which is a direct consequence of the interaction of the supercooled BEC with gravitons. The subsequent discussion in this paper will be based upon this important result that we have obtained. We shall now proceed to calculate the form of the above solution when specific incoming gravitational wave template is used and obtain the corresponding Bogoliubov coefficients.
\end{widetext}
\begin{figure}
\begin{center}
\includegraphics[scale=0.35]{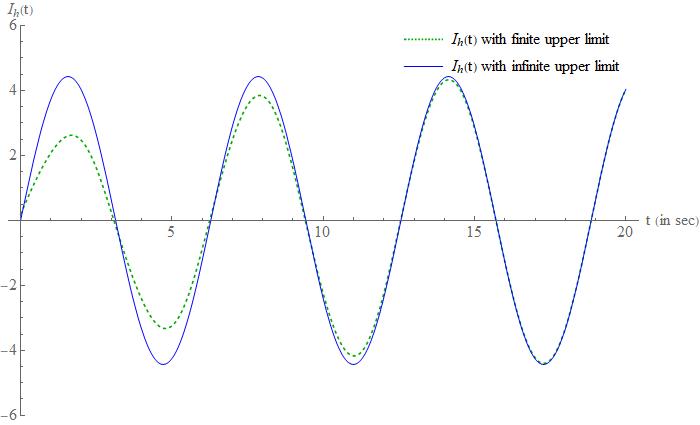}
\caption{$I_h(t)$ is plotted for $\omega_\beta=1 \text{ Hz}$, $\Omega=2\text{ Hz}$ and $\tau=10$ sec when the upper limit of integration is finite and infinite as well.}\label{F1}
\end{center}
\end{figure}
At first, it is important to note that in the last term in eq.(\ref{1.51}) the integral can be assumed to operate in a very small time limit as the interaction of BEC with a gravitational wave will occur for a very small amount of time. From eq.(\ref{1.45}), we observe that if $t$ is very small then $\Omega_m-\frac{\sin(\Omega_mt)}{t}\simeq\Omega_m-\frac{\Omega_m t}{t}=0$. Hence, for a simpler analysis, we indeed can set $\mathcal{f}_\beta(t)$ equals to zero in eq.(\ref{1.51}). The classical template of the gravitational wave can be used as $h^{\text{cl}}(t',0)=\varepsilon e^{-\frac{t^2}{\tau^2}}\sin(\Omega t)$, where $\tau$ indicates the duration of time for capturing a single measurement of the gravitational wave. From eq.(\ref{1.51}), we replace $h^{\text{cl}}(t',0)$ by the above analytical form and obtain
$I_h(t)\equiv\varepsilon\int_{-\infty}^tdt' e^{-i\omega_\beta t'}\sin(\omega_\beta(t-t'))e^{-\frac{{t'}^2}{\tau^2}}\sin(\Omega t')$. We shall make the upper limit to $\infty$ and argue that it is a quite good approximation for a such a gravitational wave template and we plot the two behaviour where the upper limit of integration is finite in one case and infinite in the other case in Figure (\ref{F1}). The quantum gravitational fluctuation term in eq.(\ref{1.51}) is very small in amplitude. Hence, we can simply assume that the noise overall can be modelled by that of the final value of the fluctuation at time $t$. Hence, we use the relation that $\int dt'\delta\hat{N}(t')g(t')\simeq\int dt'\delta\hat{N}(t)g(t')$ with $g(t)$ being a function of time. In a simple experimental setup, if the measurement time corresponding to a single BEC mode is $t=\mathcal{t}$ then the solution in eq.(\ref{1.51}) can be recast in the final form as
\begin{widetext}
\begin{equation}\label{1.52}
\begin{split}
\hat{\psi}_{\mathbf{k}_\beta}(t)&=\left(1-\frac{\mathcal{k}_0^2}{4}(1+2i\omega_\beta \mathcal{t})\delta\hat{N}(\mathcal{t})\right)e^{-i\omega_\beta t}+\left(\frac{\varepsilon \mathcal{k}_0^2}{4}\sqrt{\pi}\omega_\beta\tau\left(e^{-\frac{\tau^2}{4}(\Omega-2\omega_\beta)^2}-e^{-\frac{\tau^2}{4}(\Omega+2\omega_\beta)^2}\right)+\frac{\mathcal{k_0^2}}{4}\delta\hat{N}(\mathcal{t})\right)e^{i\omega_\beta t}\\
&=\hat{\alpha}^\beta(\mathcal{t})e^{-i\omega_\beta t}+\hat{\beta}^{\beta}(\mathcal{t})e^{i\omega_\beta t}
\end{split}
\end{equation}
where the coefficients $\hat{\alpha}^\beta(\mathcal{t})$ and $\hat{\beta}^\beta(\mathcal{t})$ are defined as
\begin{align}
\hat{\alpha}^\beta(\mathcal{t})&\equiv\alpha^\beta+\delta\hat{\alpha}^\beta(\mathcal{t})=1-\frac{{k_\beta}_x^2-{k_\beta}_y^2}{4k_\beta^2}(1+2i\omega_\beta \mathcal{t})\delta\hat{N}(\mathcal{t})\label{1.53}\\
\hat{\beta}^\beta(\mathcal{t})&\equiv\beta^\beta+\delta\hat{\beta}^\beta(\mathcal{t})=\frac{{k_\beta}_x^2-{k_\beta}_y^2}{4k_\beta^2}\sqrt{\pi}\varepsilon\omega_\beta\tau\left(e^{-\frac{\tau^2}{4}(\Omega-2\omega_\beta)^2}-e^{-\frac{\tau^2}{4}(\Omega+2\omega_\beta)^2}\right)+\frac{{k_\beta}_x^2-{k_\beta}_y^2}{4k_\beta^2}\delta\hat{N}(\mathcal{t})~.\label{1.54}
\end{align}
\end{widetext}
It is quite natural to set $\mathcal{t}=\tau$ later on but for the moment we leave them as two distinct numbers. It is straight forward to write down $\delta\hat{\alpha}^{\beta}=\mathcal{c}^{\alpha}(\mathcal{t})\delta\hat{N}(\mathcal{t})$ and $\delta\hat{\beta}^{\beta}=\mathcal{c}^{\beta}(\mathcal{t})\delta\hat{N}(\mathcal{t})$ such that $\mathcal{c}^{\alpha}(\mathcal{t})=-\frac{\tilde{\varepsilon}}{4\epsilon}(1+2i\omega_\beta t)$, and $\mathcal{c}^{\beta}(\mathcal{t})=\frac{\varepsilon}{4\varepsilon}$. Here, a new quantity is defined, $\tilde{\varepsilon}\equiv\varepsilon\frac{{k_\beta}_x^2-{k_\beta}_y^2}{k_\beta^2}$. As the background spacetime is curved, it is evident from eq.(\ref{1.52}) that $\hat{\alpha}_\beta(\mathcal{t})$ and $\hat{\beta}_\beta(\mathcal{t})$ are the Bogoliubov coefficients.

\noindent Eq.(s)(\ref{1.53},\ref{1.54}) signifies that the Bogoliubov coefficients can now be decomposed into two parts, one is the classical part which is same as that obtained in \cite{PhononBEC3}, and the another part is a fluctuation term which is indeed purely a quantum gravitational term. It is quite natural that as the background spacetime fluctuations are quantized, the Bogoliubov coefficients will not be constant numbers anymore rather the noise fluctuations will be embedded into them. This is a very important observation in this paper. Our aim is to obtain the variance in the measurement of the parameter $\varepsilon$. For the next part of our analysis, we shall make use of quantum metrology techniques to  extract signatures of the noise induced by the gravitons on the BEC. It is important to note that the metrology techniques followed in \cite{PhononBEC3}, will be a bit different in this scenario as the Bogoliubov coefficients are now having a fluctuation term. In order to inspect the quantum Fisher information term in this scenario \cite{PhononBEC3}, we finally will need to take the stochastic average of the same.
\section{Quantum metrology and the noise of gravitons}\label{S4}
\subsection{Calculating the covariance matrix for the single mode of an \textit{n} mode bosonic system}
\noindent The method that we shall follow in this literature is the covariance matrix formalism. In this subsection, we have given a pedagogical derivation of the covariance matrix of  a squeezed one mode Bose-Einstein condensate at zero temperature. For a system of $n$ bosons, the position and the conjugate momenta in terms of the ladder operators are given as
\begin{equation}\label{1.55}
\hat{x}^\beta_{j}=\sqrt{\frac{\hbar}{2m\omega_\beta}}(\hat{a}^\beta_{j}+\hat{a}^{\beta\dagger}_{j})~,~~\hat{p}^\beta_{k}=i\sqrt{\frac{m\hbar\omega_\beta}{2}}(\hat{a}^{\beta\dagger}_{k}-\hat{a}^\beta_{k})~.
\end{equation}
If one imposes the commutation relation $[\hat{a}^\beta_{k},\hat{a}^{\beta\dagger}_{k'}]=\delta_{k,k'}$, it is quite straight forward to check that $[\hat{x}^\beta_j,\hat{p}^\beta_k]=i\hbar \delta_{jk}$ with $j,k=1,\cdots,n$. Here, $\hat{a}_k^\beta$  denotes the annihilation operator corresponding to the $k$-th energy state such that $\hat{a}_k^\beta|m_k^\beta\rangle=\sqrt{m_k^\beta}|m_k^\beta-1\rangle$, and $\hat{a}_k^{\beta\dagger}$  denotes the creation operator such that $\hat{a}_k^{\beta\dagger}|m_k^\beta\rangle=\sqrt{m_k^\beta+1}|m_k^\beta+1\rangle$. The vacuum state is defined as $\hat{a}_k^\beta|0^\beta_k\rangle=0$, $\forall k$. One can now define a column vector of the form $\mathcal{R}=\left(\mathcal{u}\hat{x}^\beta_1,\frac{1}{\mathcal{v}}\hat{p}^\beta_1,\cdots,\mathcal{u}\hat{x}^\beta_n,\frac{1}{\mathcal{v}}\hat{p}^\beta_n\right)^T$
where $\mathcal{u}\equiv\sqrt{\frac{m\omega_\beta}{\hbar}}$ and $\mathcal{v}=\sqrt{m\hbar\omega_\beta}$. It is quite straight forward to check that $[\mathcal{R},\mathcal{R}^T]\equiv\mathcal{R}\mathcal{R}^T-(\mathcal{R}\mathcal{R}^T)^T=i\mathcal{O}$, where $\mathcal{O}=\bigoplus\limits_{j=1}^n i\sigma_2$ with $\sigma_2$ denoting the second Pauli spin matrix and $\bigoplus$ denoting the direct sum corresponding to the $n$-modes of the $n$-mode bosonic system. Here, $\mathcal{O}$ is $2n\times2n$ dimensional matrix. The covariant matrix $\Sigma$ in terms of the $\mathcal{R}$ column matrix reads
\begin{equation}\label{1.56}
\Sigma_{ij}=\frac{1}{2}\langle \{\mathcal{R}_i,\mathcal{R}_j\}\rangle-\langle\mathcal{R}_i\rangle\langle\mathcal{R}_i\rangle
\end{equation}
where the expectation value is taken with resect to the density matrix of the $n$ mode bosons. 
For a single mode, we can write down $\mathcal{R}^k=\left(\mathcal{u}\hat{x}^\beta_k,\frac{1}{\mathcal{v}}\hat{p}_k^\beta\right)^T$ where only the $k$-th mode is being considered. It is straight forward to check that for a single-mode vacuum state of bosons in thermal equilibrium, the density matrix reads ($k$-th bosonic mode is considered)
\begin{equation}\label{1.57}
\begin{split}
\hat{\rho}^k&=\frac{e^{-\text{\ss} \hat{H}}}{\text{tr}[e^{-\text{\ss} \hat{H}}]}\\
&=\frac{\sum\limits_{n_k^\beta=0}^\infty|n_k^\beta\rangle\langle n_k^\beta|e^{-\text{\ss} \hat{a}_{k}^{\beta\dagger} \hat{a}_{k}^\beta}}{\sum\limits_{m_k^\beta=0}^\infty\langle m_k^\beta|e^{-\text{\ss} \hat{a}_{k}^{\beta\dagger}\hat{a}_{k}^\beta}|m_k^\beta\rangle}\\
&=(1-e^{-\text{\ss}})\sum\limits_{n_k^\beta=0}^\infty|n_k^\beta\rangle\langle n_k^\beta|e^{-\text{\ss} n_k^\beta}\\
&=\frac{1}{1+\mathcal{N}}\sum\limits_{n_k^\beta=0}^\infty\left(\frac{\mathcal{N}}{1+\mathcal{N}}\right)^{m_k^\beta}|m_k^\beta\rangle\langle m_k^\beta|
\end{split}
\end{equation} 
where $\text{\ss}=\frac{1}{k_B T}$ and $\mathcal{N}=\frac{1}{e^{\text{\ss}}-1}$ with $k_B$ denoting the Boltzmann constant,  and $T$ denoting the equilibrium temperature. Using eq.(\ref{1.57}), it is straight forward to show that $\langle\mathcal{R}^k\rangle=0$. It is therefore sufficient to calculate $\frac{1}{2}\text{tr}\left[\{\mathcal{R}^k,{\mathcal{R}^k}^T\}\hat{\rho}^k\right]$ in order to obtain the covariance matrix $\Sigma^k$ corresponding to the $k$-th bosonic mode of the $n$-mode bosonic system. The anti-commutator $\{\mathcal{R}^k,{\mathcal{R}^k}^T\}$ reads
\begin{equation}\label{1.58}
\{\mathcal{R}^k,{\mathcal{R}^k}^T\}=
\begin{bmatrix}
\frac{2m\omega_\beta}{\hbar}(\hat{x}^{\beta}_k)^2&&\frac{1}{\hbar}\left(\hat{x}^{\beta}_k\hat{p}^{\beta}_k+\hat{p}^{\beta}_k\hat{x}^{\beta}_k\right)\\
\frac{1}{\hbar}\left(\hat{p}^{\beta}_k\hat{x}^{\beta}_k+\hat{x}^{\beta}_k\hat{p}^{\beta}_k\right)&&\frac{2}{m\hbar\omega_\beta}(\hat{p}_k^\beta)^2
\end{bmatrix}.
\end{equation}
Using the above matrix structure, one can obtain the final form of the covariance matrix corresponding to a single mode of an $n$-mode bosonic system as
\begin{equation}\label{1.59}
\begin{split}
\Sigma^k[\mathcal{N}(T)]&=\frac{1}{2}\text{tr}\left[\{\mathcal{R}^k,{\mathcal{R}^k}^T\}\hat{\rho}^k\right]=\frac{2\mathcal{N}+1}{2}
\begin{bmatrix}
1&&0\\
0&&1
\end{bmatrix}~.
\end{split}
\end{equation}
In the zero temperature limit, $\mathcal{N}=0$. If the entire system is at zero temperature then effectively it denotes a BEC corresponding to  the $k$-th mode of the $n$-mode bosonic system. The covariance matrix corresponding to a BEC from eq.(\ref{1.59}) can be obtained using the zero temperature limit as
\begin{equation}\label{1.60}
\begin{split}
\Sigma^k[\mathcal{N}(0)]&=\Sigma^k[0]\\&=\lim\limits_{T\rightarrow 0}\Sigma^k[\mathcal{N}(T)]\\&=\begin{bmatrix}
\frac{1}{2}&&0\\
0&&\frac{1}{2}
\end{bmatrix}~.
\end{split}
\end{equation}
In order to enhance the feedback of the BEC from the gravitational wave, the general idea is to squeeze the single mode bosons in the zero temperature limit. For the next part of our analysis, we shall drop the $k$ superscript. Under a squeezing by a parameter $r_{\text{sq.}}=re^{i\varphi}$, one can obtain the following two relations
\begin{equation}\label{1.61}
\begin{split}
\hat{S}(r)\hat{a}^\beta\hat{S}^\dagger(r)&=\hat{a}^\beta\cosh r+\hat{a}^{\beta\dagger} e^{i\varphi}\sinh r~,\\
\hat{S}(r)\hat{a}^{\beta\dagger}\hat{S}^\dagger(r)&=\hat{a}^{\beta\dagger}\cosh r+\hat{a}^\beta e^{-i\varphi}\sinh r~.
\end{split}
\end{equation}
Applying the transformations to the vector $\mathcal{R}$, we obtain
\begin{equation}\label{1.62}
\begin{split}
&\hat{S}(r)\mathcal{R}\hat{S}^\dagger(r)=\begin{pmatrix}
\mathcal{u}\hat{S}(r)\hat{x}^\beta\hat{S}^\dagger(r)\\
\frac{1}{\mathcal{v}}\hat{S}(r)\hat{p}^\beta\hat{S}^\dagger(r)
\end{pmatrix}\\
&=\begin{bmatrix}
\cosh r+\cos\varphi\sinh r&&\sin\varphi\sinh r\\
\sin\varphi\sinh r&&
\cosh r-\cos\varphi\sinh r
\end{bmatrix}\begin{pmatrix}
\mathcal{u}\hat{x}^\beta\\
\frac{1}{\mathcal{v}}\hat{p}^\beta
\end{pmatrix}
\\
&=\Xi_{\text{sq.}}(r)\mathcal{R}
\end{split}
\end{equation}
where $\Xi_{\text{sq.}}(r)$ denotes the squeezing matrix. Hence, the covariance matrix corresponding to the single-mode squeezed phonons of the BEC reads \footnote{Similar result for the squeezed covariance matrix has been reproduced earlier in \cite{PhononBEC3, continuous_variable_QI} but a slightly different result was produced where the off-diagonal elements of the matrix comes with a negative sign.}
\begin{equation}\label{1.63}
\begin{split}
&\Sigma_{\text{sq.}}[0]\\&=\Xi_{\text{sq.}}(r)~\Sigma[0]~\Xi^T_{\text{sq.}}(r)\\
&=\frac{1}{2}
\begin{bmatrix}
\cosh 2r+\cos\varphi\sinh 2r&&\sin\varphi\sinh 2r\\
\sin\varphi\sinh 2r&&\cosh 2r-\cos\varphi\sinh 2r
\end{bmatrix}~.
\end{split}
\end{equation} 
When a gravitational wave interacts with the squeezed BEC, it will transform the covariance matrix obtained in eq.(\ref{1.63}) as \cite{quantum_metrology}
\begin{equation}\label{1.64}
\begin{split}
\tilde{\Sigma}_k(\tilde{\varepsilon})=\mathcal{M}_{kk}(\tilde{\varepsilon})\Sigma_{\text{sq.}}[0]\mathcal{M}_{kk}^T(\tilde{\varepsilon})+\sum\limits_{j\neq k}\mathcal{M}_{kj}(\tilde{\varepsilon})\mathcal{M}^T_{kj}(\tilde{\varepsilon})
\end{split}
\end{equation}
where $\tilde{\varepsilon}=\varepsilon\mathcal{k}_0^2$ and $\mathcal{M}_{kj}(\tilde{\varepsilon})$ is given as \cite{PhononBEC3,quantum_metrology}
\begin{equation}\label{1.65}
\mathcal{M}_{kj}(\tilde{\varepsilon})=\begin{bmatrix}
\Re[{\alpha}^\beta_{kj}-{\beta}^\beta_{kj}]&&\Im[{\alpha}^\beta_{kj}+{\beta}^\beta_{kj}]\\
-\Im[{\alpha}^\beta_{kj}-{\beta}^\beta_{kj}]&&\Re[{\alpha}^\beta_{kj}+{\beta}^\beta_{kj}]
\end{bmatrix}
\end{equation}
with ${\alpha}^\beta_{kj}$ and $\beta^\beta_{kj}$ denoting the classical  Bogoliubov coefficients.
In our current analysis, the Bogoliubov coefficients are operators involving a small fluctuation term. The expectation value of the fluctuation term vanishes and the two point correlator has a non-vanishing contribution. Hence, the individual elements of the matrix $\mathcal{M}_{kj}(\varepsilon)$ will have an additional contribution from the noise fluctuations. The modified symplectic matrix including the effects from the noise fluctuation takes the form
\begin{equation}\label{1.66}
\begin{split}
&\doublehat{\tilde{\mathcal{M}}}_{kj}(\tilde{\varepsilon})\equiv\mathcal{M}_{kj}(\tilde{\varepsilon})+\delta\doublehat{\mathcal{M}}_{kj}(\tilde{\varepsilon})\\
&=\mathcal{M}_{kj}(\tilde{\varepsilon})+\begin{bmatrix}
\Re[\mathcal{c}^{\alpha}_{kj}-\mathcal{c}^{\beta}_{kj}]&&\Im[\mathcal{c}^{\alpha}_{kj}+\mathcal{c}^{\beta}_{kj}]\\
-\Im[\mathcal{c}^{\alpha}_{kj}-\mathcal{c}^{\beta}_{kj}]&&\Re[\mathcal{c}^{\alpha}_{kj}+\mathcal{c}^{\beta}_{kj}]
\end{bmatrix}\delta\hat{N}(\mathcal{t})
\end{split}
\end{equation} 
where we have defined a new symbol $\doublehat{A}$ which indicates a matrix $A$ with operators as its elements.
It is essential to note that $\delta\hat{N}(\mathcal{t})$ is a stochastic parameter, as a result we cannot define its eigenvalues and the non-vanishing contribution comes only from the two-point correlator of the stochastic term with the contribution being always a real number. As a result, we can call it a stochastic operator and $\delta\hat{M}_{kj}(\tilde{\varepsilon})$ carries the entire essence of the stochastic operator. It is though interesting to note that $\delta\hat{N}(\mathcal{t})$ has no well defined adjoint operator, as a result it is complementary to consider it as real operator. We can now rewrite eq.(\ref{1.64}) in terms of the modified symplectic matrices with elements including operators as
\begin{equation}\label{1.67}
\begin{split}
\doublehat{\tilde{\Sigma}}_k(\tilde{\varepsilon})=\doublehat{\tilde{\mathcal{M}}}_{kk}(\tilde{\varepsilon})\Sigma_{\text{sq.}}[0]\doublehat{\tilde{\mathcal{M}}}_{kk}^T(\tilde{\varepsilon})+\sum\limits_{j\neq k}\doublehat{\tilde{\mathcal{M}}}_{kj}(\tilde{\varepsilon})\doublehat{\tilde{\mathcal{M}}}^T_{kj}(\tilde{\varepsilon})~.
\end{split}
\end{equation} 
The Bogoliubov coefficients do not involve two different modes corresponding to the $n$-mode bosonic system. Hence, it is straight forward to express the two coefficients as $\hat{\alpha}_{kj}^\beta=\delta_{kj}\hat{\alpha}^\beta$ and $\hat{\beta}_{kj}^\beta=\delta_{kj}\hat{\beta}^\beta$. With the analytical form of $\doublehat{\tilde{\mathcal{M}}}_{kj}(\tilde{\varepsilon})$, we are now in a position to calculate the error in measurement of the parameter $\tilde{\varepsilon}$ in the next subsection.
\subsection{Quantum Fisher information}\label{IVB}
\noindent In a general scenario, classical measurement may suffice in  precise determination of parameters. In cases where quantum mechanical effects are mostly in action, it may not be possible to precisely determine the outcome of small parameter without using quantum mechanical measurement techniques. This is known as quantum metrology. 
\subsubsection{Cram\'{e}r-Rao bound and the quantum Fisher information}\label{QFSS1}
\noindent Here, we give a brief introduction to the Cram\'{e}r-Rao bound involving the classical Fisher information and the techniques used to obtain the quantum Fisher information. Consider a generalized measurement by a set of Hermitian operators $\hat{\mathcal{G}}(\zeta)$ which are nonnegative and $\int d\zeta~ \hat{\mathcal{G}}(\zeta)=\hat{\mathbb{1}}$. If the probability density for obtaining the result $\zeta$, when a parameter $\vartheta$ is given, is $p(\zeta|\vartheta)=\text{tr}[\hat{\mathcal{G}}(\zeta)\hat{\rho}(\vartheta)]$ then the classical Fisher information is defined by
\begin{equation}\label{1.68}
\begin{split}
\mathcal{I}_\vartheta&\equiv \int d\zeta~p(\zeta|\vartheta)\left[\frac{\partial \ln p(\zeta|\vartheta)}{\partial \vartheta}\right]^2\\&=\int\frac{ d\zeta}{p(\zeta|\vartheta)}\left[\frac{\partial p(\zeta|\vartheta)}{\partial \vartheta}\right]^2~.
\end{split}
\end{equation}
The minimum value in the error of the estimation of the parameter $\vartheta$ from the $\mathfrak{N}$ number of independent measurements, with the set of results $\{\zeta_1,\zeta_2,\cdots,\zeta_\mathfrak{N}\} $, is obtained using the Cram\'{e}r-Rao bound to be \cite{BraunsteinCaves}
\begin{equation}\label{1.69}
\langle (\Delta\vartheta)^2\rangle\geq \frac{1}{\mathfrak{N}\mathcal{I}_\vartheta}~.
\end{equation}
If one now considers $\vartheta$ to be a parameter corresponding to a quantum-mechanical system, then the generalized form of the classical Fisher information reads \cite{BraunsteinCaves}
\begin{equation}\label{1.70}
\mathcal{I}_\vartheta=\int d\zeta \frac{1}{\text{tr}[\hat{\mathcal{G}}(\zeta)\hat{\rho}(\vartheta)]}\text{tr}\left[\hat{\mathcal{G}}(\zeta)\frac{\partial\hat{\rho}(\vartheta)}{\partial \vartheta}\right]^2~.
\end{equation}
The quantum Fisher information, (when considering all measurements $\{\hat{\mathcal{G}}(\zeta)\}$) reads \cite{BraunsteinCaves}
\begin{equation}\label{1.71}
\mathcal{H}_\vartheta=\max\limits_{\{\hat{\mathcal{G}}(\zeta)\}}\mathcal{I}_\vartheta~.
\end{equation}
Hence, the maximum amount of information, one can extract after $\mathfrak{N}$ measurements is determined by the quantum Fisher information as
\begin{equation}\label{1.72}
\langle (\Delta\vartheta)^2\rangle\geq \frac{1}{\mathfrak{N}\mathcal{I}_\vartheta}\geq\frac{1}{\mathfrak{N}\mathcal{H}_\vartheta}~.
\end{equation}
For two states $\rho_1$ and $\rho_2$, the overlap between them is determined by the fidelity $\mathcal{F}(\rho_1,\rho_2)=\left(\text{tr}\left[\sqrt{\sqrt{\rho_1}\rho_2\sqrt{\rho}_2}\right]\right)$. One can express the quantum Fisher information in eq.(\ref{1.71}), in terms of the Fidelity between two nearby states $\rho_\vartheta$ and $\rho_{\vartheta+d\vartheta}$ as \cite{quantum_metrology}
\begin{equation}\label{1.73}
\mathcal{H}_\vartheta=\frac{8(1-\sqrt{\mathcal{F}(\rho_\vartheta,\rho_{\vartheta+d\vartheta})})}{d\vartheta^2}~.
\end{equation}
For Gaussian states, it is easier to use the covariance matrix approach than the density matrix approach. Now, the overlap between two covariance matrices $\Sigma_1$ and $\Sigma_2$ for a single mode bosonic systems reads \cite{Uhlmann2mode}
\begin{equation}\label{1.74}
\mathcal{F}(\Sigma_1,\Sigma_2)=\frac{1}{\sqrt{\Lambda+\Delta}-\sqrt{\Lambda}}
\end{equation}
where
\begin{align}
\Lambda&=\frac{1}{4}\det\left[\Sigma_1+\frac{i}{2}\mathcal{O}\right]\det\left[\Sigma_2+\frac{i}{2}\mathcal{O}\right]~,\label{1.75}\\
\Delta&=\frac{1}{4}\det\left[\Sigma_1+\Sigma_2\right]\label{1.76}~.
\end{align}
 If $\vartheta$ is a very small parameter then it is possible to perturbatively expand $\Sigma$ along with $\mathcal{F}$ and $\mathcal{H}_\vartheta$. We briefly discuss the methodology presented in \cite{quantum_metrology}. For a perturbative calculation, the initial assumption is that the Bogoliubov coefficients can be expanded upto second order in $\vartheta$ as
 \begin{equation}\label{1.77}
 \begin{split}
 \alpha_{ij}(\vartheta)&\simeq\alpha_{ij}^{(0)}+\vartheta\alpha_{ij}^{(1)}+\vartheta^2 \alpha_{ij}^{(2)}\\
 \beta_{ij}(\vartheta)&\simeq\vartheta\beta_{ij}^{(1)}+\vartheta^2 \beta_{ij}^{(2)}~. 
 \end{split} 
 \end{equation}
 The above expansion is applicable for any symplectic matrix $\mathcal{M}$ that operates on $\Sigma$ to change it to some different matrix, such that the uncertainty relation still holds true.
 As both the first and second order moments of $\mathcal{R}$ can be expanded in this manner, one can express the covariance matrix $\Sigma(\vartheta)$ upto order $\vartheta^2$ as
 \begin{equation}\label{1.78}
 \Sigma(\vartheta)\simeq\Sigma^{(0)}+\vartheta \Sigma^{(1)}+\vartheta^2\Sigma^{(2)}. 
 \end{equation}
It is straightforward to note that $\mathcal{F}(\Sigma(\vartheta),\Sigma(\vartheta))=1$ as a covariance matrix is always in a full overlap with itself. Another important criteria that is necessary to impose is, $\frac{\partial\mathcal{F}(\Sigma(\vartheta),\Sigma(\vartheta+d\vartheta))}{\partial\vartheta}\Bigr\rvert_{d\vartheta=0}=0$ \cite{BraunsteinCaves}. Using the above two conditions, one can expand $\mathcal{F}(\Sigma(\vartheta),\Sigma(\vartheta+d\vartheta))$ as
\begin{equation}\label{1.79}
\mathcal{F}(\Sigma(\vartheta),\Sigma(\vartheta+d\vartheta))=1-\frac{\mathcal{F}^{(2)}}{2}d\vartheta^2+\mathrm{O}(\vartheta d\vartheta^2+\vartheta^2d\vartheta)
\end{equation}
where $\mathcal{F}^{(2)}=\mathcal{E}^{(2)}+\mathcal{C}^{(2)}$. $\mathcal{E}^{(2)}$ is proportional to the displacement of the squeezed state and as a result it is zero. For a single mode scenario, $\mathcal{C}^{(2)}$ has the form
\begin{equation}\label{1.80}
\begin{split}
\mathcal{C}^{(2)}=&\frac{1}{2}\left(\Sigma_{11}^{(0)}\Sigma_{22}^{(2)}+\Sigma_{11}^{(2)}\Sigma_{22}^{(0)}-2\Sigma_{12}^{(0)}\Sigma_{12}^{(2)}\right)\\&+\frac{1}{8}\left(\Sigma_{11}^{(1)}\Sigma_{22}^{(1)}-2\Sigma_{12}^{(1)}\Sigma_{12}^{(1)}\right)~.
\end{split}
\end{equation}
The quantum Fisher information, in terms of $\mathcal{E}^{(2)}$ and $\mathcal{C}^{(2)}$, reads \cite{quantum_metrology}
\begin{equation}\label{1.81}
\mathcal{H}_\vartheta=4\mathcal{E}^{(2)}+4\mathcal{C}^{(2)}=4\mathcal{C}^{(2)}
\end{equation}
as $\mathcal{E}^{(2)}$ is zero for the squeezed bosonic states with no displacement parameters \cite{PhononBEC3}. In the next part of subsection (\ref{IVB}), we shall obtain the analytical extension of the quantum Fisher information when quantum-gravity effects are considered in the analysis. 
\subsubsection{Noise of gravitons and the stochastic average of the quantum Fisher information}
\begin{figure}[ht!]
\begin{center}
\includegraphics[scale=0.35]{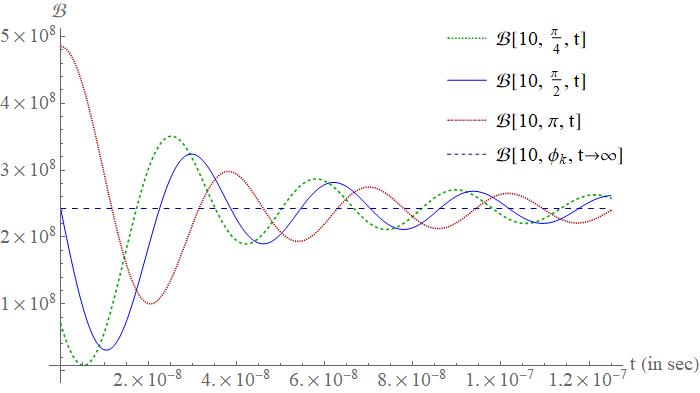}
\caption{We plot for $\phi_k=\frac{\pi}{4},\frac{\pi}{2},\pi$ and also plot the $t\rightarrow\infty$ limit of $\mathcal{B}(r_k,\phi_k,t)$ when $r_k=10$ and $\Omega_m=10^8 \text{ Hz}$.}\label{Fluctuation}
\end{center}
\end{figure}

\begin{figure}[ht!]
\begin{center}
\includegraphics[scale=0.25]{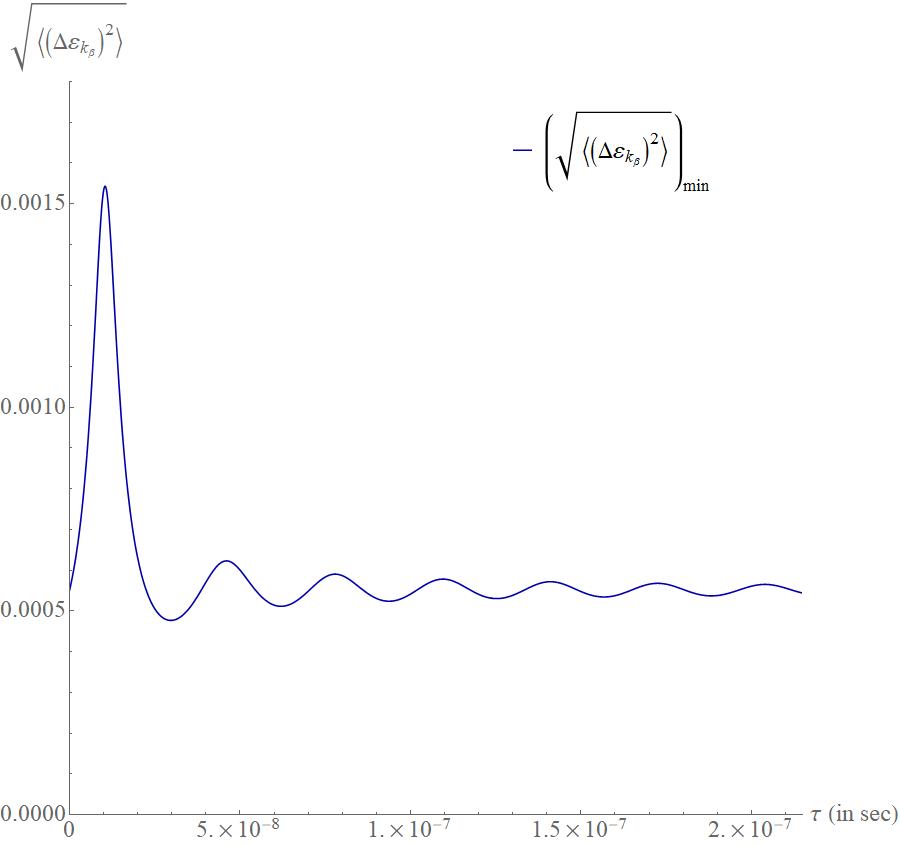}
\caption{$\sqrt{\langle(\Delta\varepsilon_{k_\beta})^2\rangle}_{\text{min.}}$ vs $\tau$ plot for the initial state of the graviton being a highly squeezed state.\label{Fish1f}}
\end{center}
\end{figure}

\begin{figure}[ht!]
\begin{center}
\includegraphics[scale=0.28]{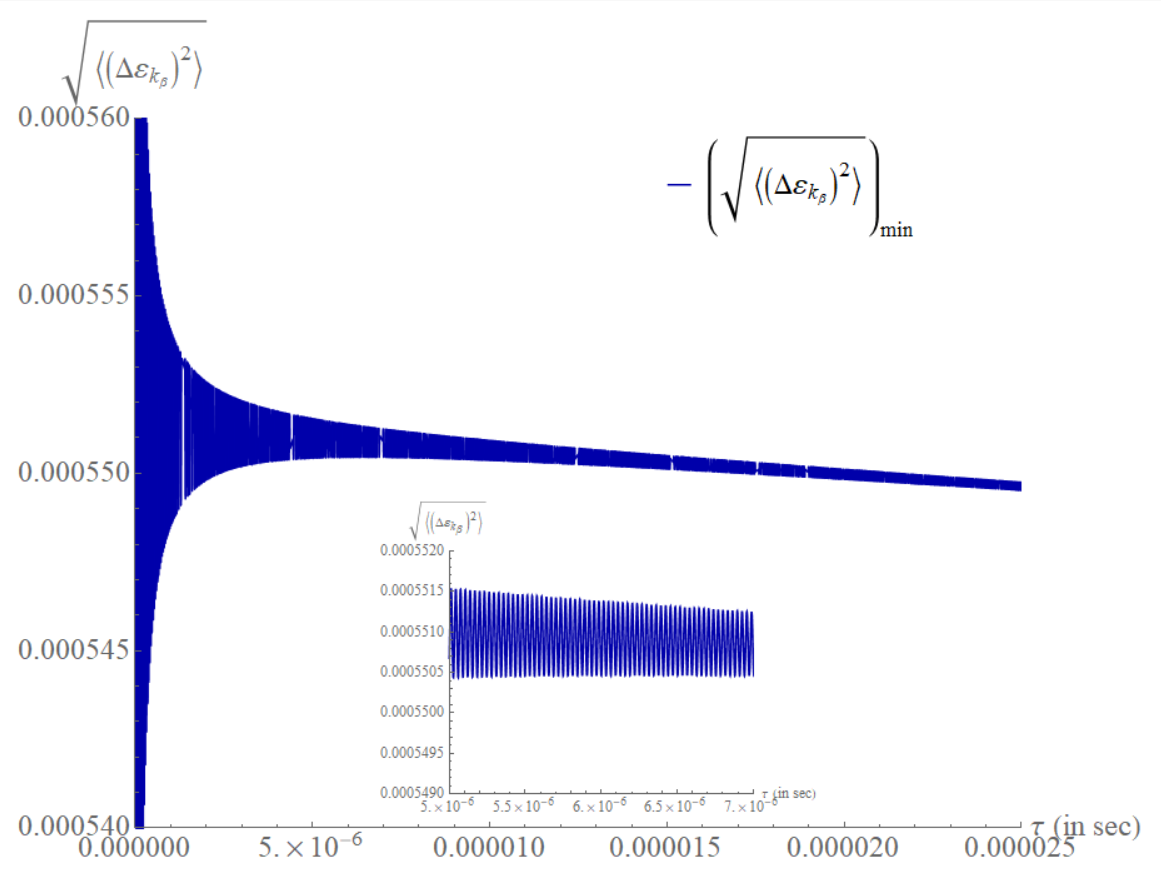}
\caption{$\sqrt{\langle(\Delta\varepsilon_{k_\beta})^2\rangle}_{\text{min.}}$ vs $\tau$ plot to observe the long time behaviour of the minimum value of the standard deviation of $\varepsilon_{k_\beta}$.\label{Fish2f}}
\end{center}
\end{figure}

\begin{figure}
\begin{center}
\includegraphics[scale=0.265]{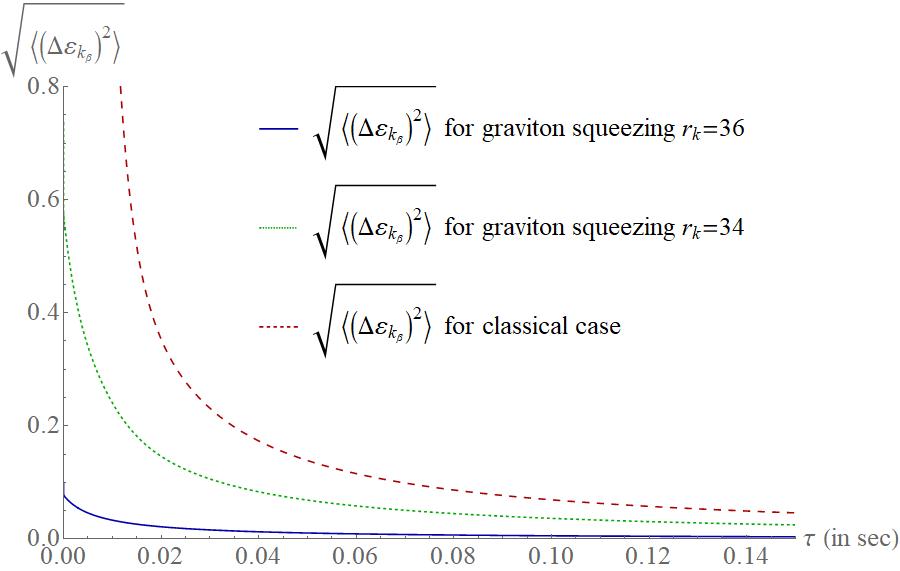}
\caption{$\sqrt{\langle(\Delta\varepsilon_{k_\beta})^2\rangle}_{\text{min.}}$ vs $\tau$ plot for squeezed gravitons with squeezing $r_k=34$ and $r_k=36$ respectively against the case of a classical gravitational wave.\label{Fish3f}}
\end{center}
\end{figure}

\begin{figure}[ht!]
\begin{center}
\includegraphics[scale=0.242]{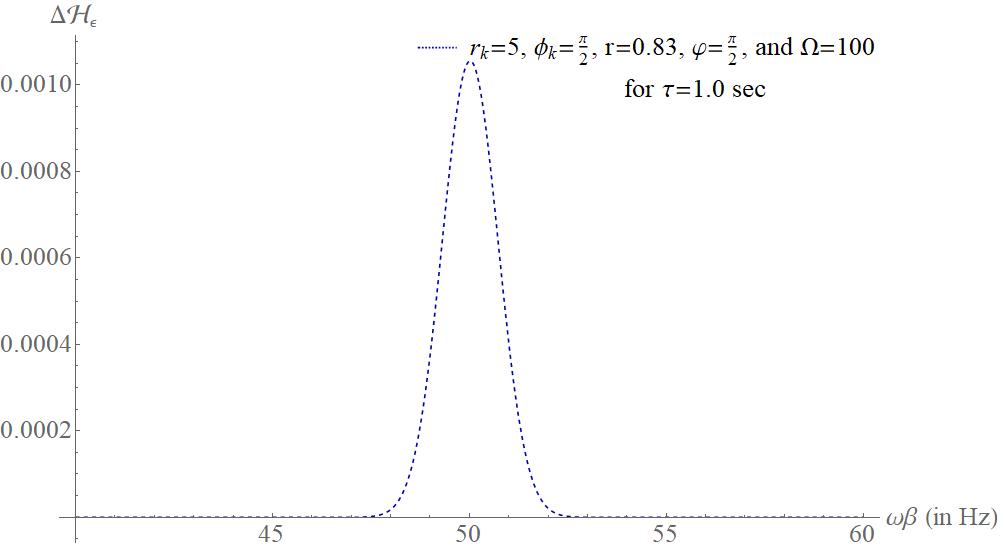}
\caption{$\Delta\mathcal{H}_\varepsilon$ vs $\omega_\beta$ plot for $\Omega_m=10^8$ Hz, $\Omega=100$ Hz, $r_k=5$, $\phi_k=\frac{\pi}{2}$, $r=0.83$, and $\varphi=\frac{\pi}{2}$. From the Figure, we can see that the peak of the standard deviation in the QGFI is observed near the resonance point $\Omega=2\omega_\beta=100$ Hz.}\label{Resonance}
\end{center}
\end{figure}

\noindent Here, we shall consider the case of the BEC interacting with gravitons. For the single mode case, the matrix $\doublehat{\tilde{M}}_{11}(\varepsilon)$ now has the form
\begin{widetext}
\begin{equation}\label{1.82}
\begin{split}
\doublehat{\tilde{M}}_{11}(\tilde{\varepsilon})=\begin{bmatrix}
1-\frac{\tilde{\varepsilon}}{2\varepsilon}\left[\delta\hat{N}(\mathcal{t})+\frac{\varepsilon\omega_\beta\tau\sqrt{\pi}}{4}\left[e^{-\frac{\tau^2}{4}(\Omega-2\omega_\beta)^2}-e^{-\frac{\tau^2}{4}(\Omega+2\omega_\beta)^2}\right]\right]&&-\frac{\tilde{\varepsilon}}{2\varepsilon}\omega_\beta t\delta\hat{N}(\mathcal{t})\\
\frac{\tilde{\varepsilon}}{2\varepsilon}\omega_\beta t\delta\hat{N}(\mathcal{t})&&1+\frac{\tilde{\varepsilon}\omega_\beta\tau\sqrt{\pi}}{4}\left[e^{-\frac{\tau^2}{4}(\Omega-2\omega_\beta)^2}-e^{-\frac{\tau^2}{4}(\Omega+2\omega_\beta)^2}\right]
\end{bmatrix}.
\end{split}
\end{equation}
Because of $\doublehat{\mathcal{M}}_{11}(\tilde{\varepsilon})$ having elemnts consisting of operators, $\tilde{\Sigma}(\varepsilon)$ from eq.(\ref{1.67}) will also have operator as its elements inspite of $\Sigma_{\text{sq.}}[0]$ having numbers as its elements. As a result $4\mathcal{C}^{(2)}$ will now have operatorial contributions in it. As a result $\mathcal{H}_{\varepsilon}$ will be operator as well.
Making use of eq.(\ref{1.82}) in eq.(\ref{1.67}), one can obtain the analytical form of $4\hat{\mathcal{C}}^{2}$ as
\begin{equation}\label{1.83}
\begin{split}
\hat{\mathcal{H}}_{\varepsilon}=&4\hat{\mathcal{C}}^{(2)}=\frac{1}{64}\pi\omega_\beta^2\tau^2\left(
e^{2\omega_\beta\Omega\tau^2}-1\right)^2
e^{-\frac{\tau^2}{2}(\Omega+2\omega_\beta)^2}\left(1+\cosh4r+(1-3\cos2\varphi)\sinh^22r\right)\\
&+\frac{\delta \hat{N}(\mathcal{t})}{32\varepsilon}\left(\sqrt{\pi}\omega_\beta\tau\right)\left(
e^{2\omega_\beta\Omega\tau^2}-1\right)
e^{-\frac{\tau^2}{4}(\Omega+2\omega_\beta)^2}\left(2\cosh^22r+(1-3\cos 2\varphi)\sinh^2 2r+6\omega_\beta \mathcal{t}\sin\varphi\sinh 4r\right)\\
&+\frac{(\delta\hat{N}(\mathcal{t}))^2}{16\varepsilon^2}\Bigr[1+\cosh 4r-\frac{\sinh^22r}{2}(3+\cos2\varphi)+\omega_\beta\mathcal{t}\bigr(\sin\varphi\left(2\sinh^22r(\cos\varphi+\omega_\beta\mathcal{t}\sin\varphi)+3\sinh 4r\right)\\&+4\omega_\beta \mathcal{t}\cosh^2 2r\bigr)\Bigr]\\
=&\mathcal{H}_\varepsilon^{(0)}+\frac{\delta\hat{N}(\mathcal{t})}{32\varepsilon}\mathcal{H}_\varepsilon^{(1)}+\frac{(\delta\hat{N}(\mathcal{t}))^2}{16\varepsilon^2}\mathcal{H}_\varepsilon^{(2)}
\end{split}
\end{equation}
where
\begin{align}
\mathcal{H}_\varepsilon^{(0)}&=\frac{1}{64}\pi\omega_\beta^2\tau^2\left(
e^{2\omega_\beta\Omega\tau^2}-1\right)^2
e^{-\frac{\tau^2}{2}(\Omega+2\omega_\beta)^2}\left(1+\cosh4r+(1-3\cos2\varphi)\sinh^22r\right)~,\label{1.84}\\
\mathcal{H}_\varepsilon^{(1)}&=\left(\sqrt{\pi}\omega_\beta\tau\right)\left(
e^{2\omega_\beta\Omega\tau^2}-1\right)
e^{-\frac{\tau^2}{4}(\Omega+2\omega_\beta)^2}\left(2\cosh^22r+(1-3\cos 2\varphi)\sinh^2 2r+6\omega_\beta \mathcal{t}\sin\varphi\sinh 4r\right)~,\label{1.85}\\
\mathcal{H}_\varepsilon^{(2)}&=1+\cosh 4r-\frac{\sinh^22r}{2}(3+\cos2\varphi)+\omega_\beta\mathcal{t}\bigr(\sin\varphi\left(2\sinh^22r(\cos\varphi+\omega_\beta\mathcal{t}\sin\varphi)+3\sinh 4r\right)+4\omega_\beta \mathcal{t}\cosh^2 2r\bigr)\label{1.86}
\end{align}
\end{widetext}
where $\mathcal{H}_\varepsilon^{(0)}$ is the quantum Fisher information for the classical contribution of the gravitational wave and is exactly similar to the result obtained in \cite{PhononBEC3}. The other two terms determine the quantum gravitational contribution to the quantum Fisher information.
The quantum Fisher information operator (or the graviton-noise induced Fisher information) is not entirely a measurable quantity now. Instead of calling it a Fisher information operator, it is better to call it a quantum gravitational Fisher information (QGFI). The straightforward way is to take a stochastic average of the quantity with respect to the graviton state as
\begin{equation}\label{1.87}
\begin{split}
\llangle \hat{\mathcal{H}}_{\varepsilon}\rrangle=\mathcal{H}_\varepsilon^{(0)}+\frac{\llangle\{\delta\hat{N}(\mathcal{t}),\delta\hat{N}(\mathcal{t})\}\rrangle}{32\varepsilon^2}\mathcal{H}_\varepsilon^{(2)}
\end{split}
\end{equation}
where the second term from eq.(\ref{1.83}) vanishes because the one point correlator of the noise operator, $\llangle\delta \hat{N}(t)\rrangle$ vanishes. Our next aim is to obtain the analytical form of $\llangle\{\delta\hat{N}(\mathcal{t}),\delta\hat{N}(\mathcal{t})\}\rrangle$ for the gravitons initially being in a squeezed state. The two point noise correlator for an arbitrary state of the graviton reads
\begin{equation}\label{1.88}
\begin{split}
&\llangle \{\delta\hat{N}_{ij}(t),\delta\hat{N}_{lm}(t')\}\rrangle\\&=\frac{4\kappa^2}{V}\sum\limits_{\mathbf{k},\mathbf{k}'}\sum\limits_{s,s'}\epsilon^s_{ij}(\mathbf{k})\epsilon^{s'}_{lm}(\mathbf{k}')\times\llangle\{\delta\hat{h}^s_I(\mathbf{k},t),\delta\hat{h}^{s'}_I(\mathbf{k}',t')\}\rrangle\\
&=\frac{2\kappa^2}{5\pi^2}(\delta_{il}\delta_{jm}+\delta_{im}\delta_{jl}-\frac{2}{3}\delta_{ij}\delta_{lm})\smash{\int_0^{\Omega_m}}dkk^2\mathcal{Q}_{\delta h}(t,t',\mathbf{k})
\end{split}
\end{equation}
where in order to obtain the final line of the above equation, we have made use of the identification between the summation over all possible modes to an integral over a continuous variable and another definition is used $\llangle\{\delta\hat{h}^s_I(\mathbf{k},t),\delta\hat{h}^{s'}_I(\mathbf{k}',t')\}\rrangle=\delta_{ss'}\delta_{\mathbf{k}+\mathbf{k}',0}\mathcal{Q}_{\delta h}(t,t',\mathbf{k})$ \cite{KannoSodaTokuda}.
For the graviton initially being in a squeezed state with squeezing parameter $r_k^{\text{sq.}}=r_ke^{i\phi_k}$, $\mathcal{Q}_{\delta h}(t,t',\mathbf{k})$ takes the form\footnote{For a detailed discusssion on the graviton state with squeezing, see Appendix (\ref{AppendixA}).}
\begin{equation}\label{1.89}
\begin{split}
\mathcal{Q}_{\delta h}(t,t',\mathbf{k})&=\frac{1}{k}\left(\cos\left(k(t-t')\right)\cosh 2r_k\right.\\&\left.-\cos\left(k(t+t')-\phi_k\right)\sinh 2r_k\right)~.
\end{split}
\end{equation} 
Using eq.(\ref{1.89}), one can obtain the two point correlator for $t=t'$ as
\begin{equation}\label{1.90}
\begin{split}
&\llangle \{\delta\hat{N}_{ij}(t),\delta\hat{N}_{lm}(t)\}\rrangle=\frac{\kappa^2\Omega_m^2}{5\pi^2}\Bigr(\delta_{il}\delta_{jm}+\delta_{im}\delta_{jl}\\&-\frac{2}{3}\delta_{ij}\delta_{lm}\Bigr)\biggr(\cosh 2r_k
+\frac{1}{2\Omega_m^2t^2}\sinh 2r_k\bigr(\cos\phi_k\\&-\cos(2\Omega_mt-\phi_k)-2\Omega_mt\sin(2\Omega_mt-\phi_k)\bigr)\biggr)~.
\end{split}
\end{equation}
The $\llangle\{\delta\hat{N}(\mathcal{t}),\delta\hat{N}(\mathcal{t})\}\rrangle$ correlator can now be obtained as
\begin{equation}\label{1.91}
\begin{split}
&\llangle\{\delta\hat{N}(\mathcal{t}),\delta\hat{N}(\mathcal{t})\}\rrangle\equiv\llangle\{\delta\hat{N}_{11}(\mathcal{t}),\delta\hat{N}_{11}(\mathcal{t})\}\rrangle\\
&=\frac{4\kappa^2\Omega_m^2}{15\pi^2}\mathcal{B}(r_k,\phi_k,\mathcal{t})
\end{split}
\end{equation} 
where the time dependent part of the two-point noise correlator reads
\begin{equation}\label{1.92}
\begin{split}
&\mathcal{B}(r_k,\phi_k,\mathcal{t})=\cosh 2r_k
+\frac{1}{2\Omega_m^2\mathcal{t}^2}\sinh 2r_k\bigr(\cos\phi_k\\&-\cos(2\Omega_m\mathcal{t}-\phi_k)-2\Omega_m\mathcal{t}\sin(2\Omega_m\mathcal{t}-\phi_k)\bigr)~.
\end{split}
\end{equation}
The squeezing in the initial graviton states can be very high, even of the order of $r_k\sim 21$ for primordial gravitational wave generated during the inflationary period \cite{KannoSodaTokuda}. For $\mathcal{t}\rightarrow 0$ limit $\mathcal{B}(r_k,\phi_k,\mathcal{t})$ has the value $\lim\limits_{\mathcal{t}\rightarrow0}\mathcal{B}(r_k,\phi_k,\mathcal{t})=\cosh 2r_k-\cos\phi_k\sinh 2r_k$ which never vanishes irrespective of any values of $\phi_k$. It is although important to note that for $\phi_k=\pi$, $\mathcal{B}(r_k,\phi_k,0)$ becomes maximum and for $\phi_k=0$, it becomes minimum. For a grand unified theory gravitational wave, the cut-off frequency is at around $\Omega_m\sim10^8 \text{ Hz}$. In order to observe the nature of the function $\mathcal{B}$ with respect to time, we use $\Omega_m\sim 10^8\text{Hz}$, $r_k\sim 10$, and $\phi_k=\{\frac{\pi}{4},\frac{\pi}{2},\pi\}$ and plot $\mathcal{B}(r_k,\phi_k,t)$ against $t$ for the above values in Fig.(\ref{Fluctuation}). It is important to note that irrespective of the $\phi_k$, in the $t\rightarrow\infty$ limit it always fluctuates about the same value $\mathcal{B}(10,\phi_k,\infty)=\cosh 20\simeq2.426\times10^8$. It is easy to observe from Fig.(\ref{Fluctuation}) as well as the infinite-time limit that for $\phi_k=\frac{\pi}{2}$, $\mathcal{B}(r_k,\frac{\pi}{2},0)=\mathcal{B}(r_k,\frac{\pi}{2},\infty)$. 
The stochastic average of the quantum Fisher information from eq.(\ref{1.87}), after using eq.(\ref{1.91}) takes the form (with proper dimensional reconstruction)
\begin{equation}\label{1.93}
\begin{split}
\llangle\hat{\mathcal{H}}_\varepsilon\rrangle=\mathcal{H}_\varepsilon^{(0)}+\frac{\hbar G\Omega_m^2}{15 \pi\varepsilon^2c^5}\mathcal{B}(r_k,\phi_k,\mathcal{t})\mathcal{H}_{\varepsilon}^{(2)}
\end{split}
\end{equation}
where $\kappa^2=\frac{8\pi\hbar G}{c^3}$.  It is now reasonable to replace $\mathcal{t}$ by $\tau$ in eq.(\ref{1.93}) as the total observation time will be equal to the single mode measurement time by the BEC. Under this condition, we can recast eq.(\ref{1.93}) as (for $\varphi=\frac{\pi}{2}$)
\begin{widetext}
\begin{equation}\label{1.94}
\begin{split}
\llangle\hat{\mathcal{H}}_\varepsilon\rrangle&=\mathcal{H}_\varepsilon^{(0)}+\frac{l_p^2\Omega_m^2}{15 \pi\varepsilon^2c^2}\mathcal{B}(r_k,\phi_k,\tau)\mathcal{H}_{\varepsilon}^{(2)}\\
&=\frac{1}{64}\pi\omega_\beta^2\tau^2\left(
e^{2\omega_\beta\Omega\tau^2}-1\right)^2
e^{-\frac{\tau^2}{2}(\Omega+2\omega_\beta)^2}\left(1+\cosh4r+4\sinh^22r\right)+\frac{l_p^2\Omega_m^2}{30 \pi\varepsilon^2c^2}\bigr(3+2\omega_\beta^2\tau^2+\cosh 4r+6\omega_\beta\tau\\&\times\sinh4r
+6\omega_\beta^2\tau^2\cosh4r\bigr)\left(\cosh 2r_k
+\frac{1}{2\Omega_m^2\tau^2}\sinh 2r_k\left(\cos\phi_k-\cos(2\Omega_m\tau-\phi_k)-2\Omega_m\tau
\sin(2\Omega_m\tau-\phi_k)\right)\right)~.
\end{split}
\end{equation}
Eq.(\ref{1.94}) is one of the main results in our paper. Setting the squeezing angle to a certain value (here, $\varphi=\frac{\pi}{2}$) is possible and has been experimentally done \cite{Chelkowski,Johnsson}.
\end{widetext}
Using eq.(\ref{1.72}), we can write down the inequality in a quantum gravitational setup as
\begin{equation}\label{1.95}
\begin{split}
\langle (\Delta\tilde{\varepsilon})^2\rangle
\geq\frac{1}{\mathfrak{N}\llangle \hat{\mathcal{H}}_\varepsilon\rrangle}~.
\end{split}
\end{equation}
Here, we are considering single mode Bose-Einstein condensate only. 
\noindent As a result, we can still write down the following relation
\begin{equation}\label{1.96}
\begin{split}
\langle (\Delta\tilde{\varepsilon}_{k_\beta})^2\rangle=\left(
\frac{k_{\beta_x}^2-k_{\beta_y}^2}{k_\beta^2}\right)^2\langle (\Delta\varepsilon_{k_\beta})^2\rangle~.
\end{split}
\end{equation}
It is now possible to express $k_{\beta_x}$ and $k_{\beta_y}$ in the spherical polar coordinates as $k_{\beta_x}=k_\beta\sin\theta_\beta\cos\phi_\beta$, and $k_{\beta_y}=k_\beta\sin\theta_\beta\sin\phi_\beta$. Using the spherical polar representation, we obtain $\langle (\Delta\tilde{\varepsilon}_{k_\beta})^2\rangle=\sin^4\theta_\beta\cos^22\phi_\beta\langle (\Delta\varepsilon_{k_\beta})^2\rangle~.
$ Doing an integral over the first quadrant of the spherical coordinate basis for all single mode bosonic state of the BEC with momentum $k_\beta$, we obtain $\int d\Omega_\beta=\int_0^{\frac{\pi}{2}}d\theta_\beta\sin^5\theta_\beta\int_0^{\frac{\pi}{2}}d\phi_\beta\cos^{2}2\phi_\beta=\frac{2\pi}{15}$. In particular it is always possible to construct a Bose-Einstein condensate such that the ground state of the bosonic system consists of only single mode bosons which is $k_\beta$ in the current case. Eq.(\ref{1.95}) can then be recast in the following form
\begin{equation}\label{1.97}
\langle (\Delta\varepsilon_{k_\beta})^2\rangle\geq\frac{15}{2\pi\mathfrak{N}\llangle\hat{\mathcal{H}}_\varepsilon\rrangle}. 
\end{equation}
If the time taken for $\mathfrak{N}$ multiple measuerments of the BEC state is $\mathfrak{t}$ then $\mathfrak{t}\simeq\mathfrak{N}\tau$. At first, we consider a single measurement of the  BEC state which indicates $\mathfrak{N}=1$. From eq.(\ref{1.97}) it is evident that $
\sqrt{\langle(\Delta\varepsilon_{k_\beta})^2\rangle}$ has the minimum value at $\sqrt{\langle(\Delta\varepsilon_{k_\beta})^2\rangle}_{\text{min.}}=\sqrt{\frac{15}{2\pi\llangle\hat{\mathcal{H}}_\varepsilon\rrangle}}$. We shall now plot the minimum value of the standard deviation in the amplitude $\varepsilon$ for a single phonon mode of the BEC against the observation time $
\tau$ in Fig.(\ref{Fish1f}). In order to plot the parameter values used are $r=0.82,\varphi=\frac{\pi}{2},r_k=42,\phi_k=\frac{\pi}{2},\Omega=100 \text{ Hz}$, and $\omega_\beta=50\text{ Hz}$.
From Fig.(\ref{Fish1f}), it is straightforward to observe that the minimum standard deviation in the measurement of $\varepsilon_{k_\beta}$ corresponding to a single mode of the BEC is not very high indicating a finite chance of observation of the graviton. It is although very important to note that with a decrease in the squeezing of the graviton state, the minimum value in the measurement of the standard deviation of the gravitational wave amplitude becomes very high indicating a non detectability of such a scenario.  We can look for the long time behaviour of the $\sqrt{\langle(\Delta\varepsilon_{k_\beta})^2\rangle}_{\text{min.}}$ in Fig.(\ref{Fish2f}) (with same parameters as used to plot Fig.(\ref{Fish1f})).
We can observe from Fig.(\ref{Fish2f}) that $\sqrt{\langle(\Delta\varepsilon_{k_\beta})^2\rangle}_{\text{min.}}$ decreases with increased time of the single measurement of the gravitational wave indicating a higher chance in detection of the gravity wave. It is important to note that the graviton signature precisely lies in the detection of resonance pulses in the single mode BEC initially. We can also check this analytically. It is important to note that in the $l_p\rightarrow 0$ limit, the contribution from the linearized quantum gravity theory vanishes in eq.(\ref{1.94}) reducing it to the result obtained in \cite{PhononBEC3} which is expected. In this $l_p\rightarrow0$ limit, if $\tau$ is set to zero, the inequality in eq.(\ref{1.97}) becomes $(\Delta\varepsilon_{k_\beta})^2\rangle\geq\infty$
implying that no gravitational wave will be detected in such a semiclassical scenario. The result for the quantum gravitational perspective becomes highly bizzarre. We observe that 
\begin{equation}\label{1.98}
\lim\limits_{\tau\rightarrow0}\llangle\hat{\mathcal{H}}_\varepsilon\rrangle=\frac{l_p^2\Omega_m^2}{15 \pi\varepsilon^2c^2}\cosh^22r(\cosh 2r_k-\cos\phi_k\sinh 2r_k).
\end{equation}
For, no squeezing of the initial gravitational wave state ($r_k=0$), we obtain $\lim\limits_{\tau\rightarrow0}\llangle\hat{\mathcal{H}}_\varepsilon\rrangle=\frac{l_p^2\Omega_m^2}{15 \pi\varepsilon^2c^2}\sim10^{-31}$ ($\Omega_m\sim10^8$ and $\varepsilon\sim 10^{-21}$). This indicates, $\sqrt{\langle(\Delta \varepsilon)^2\rangle}\geq \frac{15\varepsilon c}{l_p \Omega_m \sqrt{2}}\sim 10^{16}$. Such a high minimum value of the  $\sqrt{\langle(\Delta \varepsilon)^2\rangle}$ parameter indicates a very low sensitivity of the BEC towards the gravitational wave initially implying an impossible detection scenario. But things quickly change for a non-vanishing squeezing of the initial graviton state. Suppose that the initial squeezing angle is $\phi_k=\frac{\pi}{2}$.  For $\langle(\Delta \varepsilon)^2\rangle\simeq 1$, the squeezing will be as high as $r_k\simeq 35$. For a grand unified theory inflation $r_k\simeq 42$ \cite{KannoSodaTokuda}, $\langle(\Delta \varepsilon)^2\rangle\simeq 10^{-6}$. This is very anti-intuitive in a sense that there is a finite possiblity of detecting primordial gravitational waves from the inflationary time without a proper time interval of the detector to interact with the gravity wave. In a linearized quantum gravity model, this is not at all very unphysical as there is a linearized perturbation field around the BEC even when $\tau=0$. This will indicate the existence of a gravitons in future generation of BEC based gravitational wave detection scenario. It is also a possibility that the BEC itself starts behaving as a gravitating object which may be a very vague assumption and would not be explored in details in this literature. We leave this investigation for a future work. Finally, we plot $\sqrt{\langle(\Delta\varepsilon_{k_\beta})^2\rangle}_{\text{min.}}$ vs $\tau$ for various squeezing of the graviton state against the classical case when no gravitons are present in Fig.(\ref{Fish3f}). For the quantum gravitational case, we have used the initial squeezing angle of the graviton to be equal to $\frac{\pi}{2}$. For the BEC state we have used a squeezing of $1.4$ and a squeezing angle of $\frac{\pi}{2}$ along with the mode frequency is considered to be at $\omega_\beta=50$ Hz. As a result it will be more sensitive for incoming gravitational wave with frequency $100$ Hz. It is though important to note that most primordial gravitational waves are supposed to have a frequency in the $10^{-1}-10$ Hz range. 
We can observe very important features from Fig.(\ref{Fish3f}). We observe that with the increase in squeezing $\sqrt{\langle(\Delta\varepsilon_{k_\beta})^2\rangle}_{\text{min.}}$ becomes smaller for short measurement periods. It can be seen from the classical gravitational wave case that $\sqrt{\langle(\Delta\varepsilon_{k_\beta})^2\rangle}_{\text{min.}}$ diverges near the $\tau\rightarrow0$ limit indicating a non-detection of any gravitational wave. The above result henceforth confirms that for a quantum gravity scenario, the minimum value of the standard deviation of the gravity wave amplitude parameter never becomes zero and can be arbitrarily reduced with squeezed graviton state indicating a higher chance at proving the existence of gravitons. The next thing that is important to observe is if there is a standard deviation present in the QGFI. The standard deviation in the QGFI, reads
\begin{equation}\label{1.99}
\begin{split}
(\Delta \mathcal{H}_{\varepsilon})^2&=\llangle (\hat{\mathcal{H}}_\varepsilon-\llangle \hat{\mathcal{H}}_\varepsilon\rrangle)^2\rrangle~.
\end{split}
\end{equation}
It is quite straightforward to understand that all odd order correlators will vanish. Hence, the surving contributions will come from even order correlators. As the QGFI is calculated upto the second order correlator, we shall restrict ourselves to second order in the noise correlators only. The result in eq.(\ref{1.99}) reads
\begin{equation}\label{1.100}
\begin{split}
(\Delta \mathcal{H}_{\varepsilon})^2&\simeq \frac{\bigr(\mathcal{H}_\varepsilon^{(1)}\bigr)^2}{2048\varepsilon^2}\llangle\{\delta\hat{N}(\tau),\delta\hat{N}(\tau)\}\rrangle\\
&\simeq\frac{\hbar G\Omega_m^2\bigr(\mathcal{H}_\varepsilon^{(1)}\bigr)^2}{960 \pi \varepsilon^2c^5}\mathcal{B}(r_k,\phi_k,\tau)~.
\end{split}
\end{equation}
\begin{widetext}
The standard deviation in the QGFI can be expressed in an extended form given as ($\varphi=\frac{\pi}{2}$)
\begin{equation}
\begin{split}
(\Delta\mathcal{H}_\varepsilon)^2&=\frac{l_p^2\omega_\beta^2\Omega_m^2\tau^2}{960\varepsilon^2c^2}\bigr(e^{-\frac{\tau^2}{4}(\Omega-2\omega_\beta)^2}-e^{-\frac{\tau^2}{4}(\Omega+2\omega_\beta)^2}\bigr)^2\left(2\cosh^22r+4\sinh^2 2r+6\omega_\beta \tau\sinh 4r\right)^2\mathcal{B}(r_k,\phi_k,\tau).
\end{split}
\end{equation}
\end{widetext}
We shall now look at the behaviour of the standard deviation in the QGFI around resonance. For a finite measurement of $\tau=1$ sec, with $r_k=5,\phi_k=\frac{\pi}{2},r=0.83$, and $\varphi=\frac{\pi}{2}$, we plot $\Delta \mathcal{H}_\varepsilon$ against the phonon frequency $\omega_\beta$ for an incoming gravitational wave with frequency 0.1 kHz in Fig. (\ref{Resonance}).
We find out from Fig.(\ref{Resonance}) that the standard deviation in the stochastic QGFI will be maximum for the resonance condition which is at $\Omega=2\omega_\beta=100$ Hz. Our next aim is to obtain the QGFI when the induced noise parameter has a similar decay factor as for the classical gravitational wave case. What is important to note that the analysis without any decay factor is much more realistic than the one that we are going to investigate as any kind of classical decay should not affect the quantum gravitational influences. At resonance point, with high enough squeezing from the gravitons, it is possible to enhance the standard deviation of the QGFI to such an extent that it becomes measurable. Another important aspect can be obtained by varying the total measurement time-scale $\tau$ for such a scenario.
\begin{figure}[ht!]
\begin{center}
\includegraphics[scale=0.242]{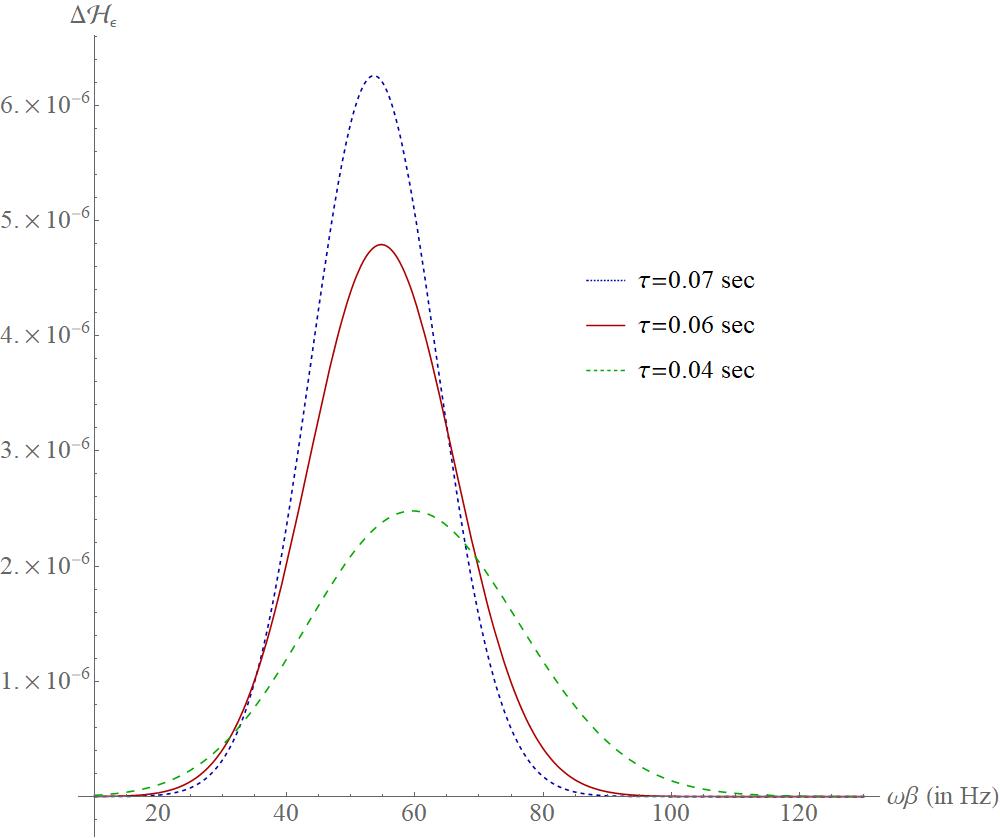}
\caption{$\Delta\mathcal{H}_\varepsilon$ vs $\omega_\beta$ plot for $\Omega_m=10^8$ Hz, $\Omega=100$ Hz, $r_k=5$, $\phi_k=\frac{\pi}{2}$, $r=0.83$, and $\varphi=\frac{\pi}{2}$. We have plotted $\Delta\mathcal{H}_\varepsilon$ for different values of $\tau$.}\label{Resonance2}
\end{center}
\end{figure}
It is very important to observe that the standard deviation in the QGFI becomes very small with a small measurement time and gets shifted towards the right for lower values of time $\tau$. This indicates that in order to detect a standard deviation in the QGFI, one needs to measure it for a longer time period.
\section{Quantum Fisher information for a decaying noise function}\label{S5}
\noindent We recall the final solution of the time dependent part of the pseudo-Goldstone boson in eq.(\ref{1.51}) and now use the form of the general noise fluctuation term at time $t'$ as $\delta\hat{N}(t')=\cos\Omega t' e^{-\frac{{t'}^2}{\tau^2}}\delta\hat{N}(t). $
The solution of the time dependent part of the pseudo-Goldstone boson takes the form for the above noise term as
\begin{equation}\label{1.102}
\hat{\psi}_{\mathbf{k}_\beta}(t)=\hat{\alpha}^\beta_\delta e^{-i\omega_\beta t}+\hat{\beta}^\beta_\delta e^{i\omega_\beta t}
\end{equation}
where the stochastic Bogoliubov coefficients take the form
\begin{align}
\hat{\alpha}^\beta_\delta&=1-\frac{2\tilde{\varepsilon}\sqrt{\pi}}{\varepsilon}\omega_\beta\tau e^{-\frac{\Omega^2\tau^2}{4}}\delta\hat{N}(\tau)\label{1.103}~,\\
\hat{\beta}^\beta_\delta&=\frac{\tilde{\varepsilon}}{4}\sqrt{\pi}\omega_\beta\tau\left(e^{-\frac{\tau^2}{4}(\Omega-2\omega_\beta)^2}-e^{-\frac{\tau^2}{4}(\Omega+2\omega_\beta)^2}\right)\nonumber\\&+\frac{\tilde{\varepsilon}}{\varepsilon}\sqrt{\pi}\omega_\beta\tau\delta\hat{N}(\tau)\left(e^{-\frac{\tau^2}{4}(\Omega-2\omega_\beta)^2}-e^{-\frac{\tau^2}{4}(\Omega+2\omega_\beta)^2}\right)~.\label{1.104}
\end{align}
One can therefore obtain the stochastic average for the QGFI as follows
\begin{widetext}
\begin{equation}\label{1.105}
\begin{split}
\llangle \hat{\mathcal{H}}_\varepsilon\rrangle=&\mathcal{H}_\varepsilon^{(0)}+\frac{\kappa^2\omega_\beta^2\Omega_m^2\tau^2}{15\varepsilon^2c^2}e^{-\frac{\tau^2}{2}(\Omega+2\omega_\beta)^2}\biggr[\left[e^{2\omega_\beta\Omega\tau^2}+1\right]^2
\times(1+\cosh 4r+4\sinh^22r)+4(7-\cosh4r)e^{2\omega_\beta^2\tau^2}\\\times&e^{2\omega_\beta\Omega\tau^2}\biggr]\mathcal{B}(r_k,\phi_k,\tau).
\end{split}
\end{equation}
\end{widetext}
Unlike the previous case, the stochastic average of the QGFI vanishes initially which is solely due to the decaying nature of the noise parameter. We shall now sum over all possible modes of the BEC. Before proceeding further, it is important to note that the BEC was initially quantized within a box of size $V_\beta=L_\beta^3$ and as a result $k_\beta=\frac{\pi n_\beta}{L_\beta}$.
Eq.(\ref{1.96}) can now be recast into the following form

\begin{equation}\label{1.106}
\begin{split}
\langle (\Delta\tilde{\varepsilon})^2\rangle&=
\sum_{\mathbf{k}_\beta}\left(
\frac{k_{\beta_x}^2-k_{\beta_y}^2}{k_\beta^2}\right)^2\langle (\Delta\varepsilon)^2\rangle\\
\implies\frac{1}{\langle (\Delta\varepsilon)^2\rangle}&=\sum_{\mathbf{k}_\beta}\left(
\frac{k_{\beta_x}^2-k_{\beta_y}^2}{k_\beta^2}\right)^2\mathfrak{N}\llangle\hat{\mathcal{H}}_\varepsilon\rrangle
\end{split}
\end{equation}
\begin{widetext}
\noindent where in the last line of the above equation, we have made use of the equality condition from eq.(\ref{1.95}) in a quantum gravitational setup. Converting the above sum over all possible modes to an integration over all possible modes and defining $\mathfrak{H}(n_\beta)\equiv \llangle\hat{\mathcal{H}}_\varepsilon\rrangle$, we obtain
\begin{equation}\label{1.107}
\begin{split}
\frac{1}{\langle (\Delta\varepsilon)^2\rangle}=\frac{2\pi}{15}\int_0^\infty dn_\beta~n_\beta^2\mathfrak{H}(n_\beta)~.
\end{split}
\end{equation}
The inequality in this case can be written as
\begin{equation}\label{1.108}
\begin{split}
\frac{1}{\mathfrak{N}\langle (\Delta\varepsilon)^2\rangle}&\leq \frac{c_s^2 \pi^4 \tau^2}{480L_\beta^2}\mathfrak{r}_1\int_0^\infty dn_\beta~ n_\beta^4 e^{-\frac{(2\pi c_sn_\beta+\Omega L_\beta)^2\tau^2}{2L_\beta^2}} \left(e^{\frac{2\pi c_s n_\beta \Omega \tau^2}{L_\beta}}-1\right)^2+\frac{2\kappa^2\pi^2c_s^2\Omega_m^2\tau^2}{15 \varepsilon^2L_\beta^2}\mathcal{B}(r_k,\phi_k,\tau)\\&\times \int_0^\infty dn_\beta~n_\beta^4 e^{-\frac{(2\pi c_sn_\beta+\Omega L_\beta)^2\tau^2}{2L_\beta^2}}\biggr[\mathfrak{r}_1\Bigr(e^{\frac{2\pi c_s n_\beta\Omega\tau^2}{L_\beta}}+1\Bigr)^2+\mathfrak{r}_2e^{\frac{2\pi c_s n_\beta \tau^2\left(\Omega+\frac{\pi c_sn_\beta}{L_\beta}\right)}{L_\beta}}\biggr]\\
&= \frac{c_s^2 \pi^4 \tau^2}{480L_\beta^2}\mathfrak{r}_1\mathfrak{I}_1+\frac{2\kappa^2\pi^2c_s^2\Omega_m^2\tau^2}{15 \varepsilon^2L_\beta^2}\mathcal{B}(r_k,\phi_k,\tau)\mathfrak{I}_2
\end{split}
\end{equation}
\end{widetext}
where
\begin{equation}\label{1.109}
\mathfrak{r}_1=1+\cosh 4r+4\sinh^2 2r,~\mathfrak{r}_2=\frac{1}{2}(7-\cosh 4r)
\end{equation}
with $\mathfrak{I}_1$ and $\mathfrak{I}_2$ denoting the first and second integral in eq.(\ref{1.108}). We shall at first explicitly investigate $\mathfrak{I}_2$ as
\begin{equation}\label{1.110}
\begin{split}
\mathfrak{I}_2=&\mathfrak{r}_1\int_0^\infty dn_\beta~ n_\beta^4 e^{-\tau^2\left[\frac{2\pi^2 c_s^2n_\beta^2}{L_\beta^2}+\frac{\Omega^2}{2}\right]}\cosh^2\left[\frac{\pi c_s n_\beta \Omega \tau^2}{L_\beta}\right]\\
+&\mathfrak{r}_2 e^{-\frac{\Omega^2\tau^2}{2}}\int_0^\infty dn_\beta~ n_\beta^4~.
\end{split}
\end{equation}
The second part of the above integral is divergent. Hence, a way out is to set the squeezing $r$ of the phonons at such a value that $\mathfrak{r}_2$ vanishes effectively. It is straight forward to check that for $r=\frac{1}{4}\cosh^{-1}(7)\simeq 0.66$, $\mathfrak{r}_2$ vanishes. It is possible to control the squeezing of the phonons. Squeezing of phonons as high as $r\simeq 0.83$ ($7.2$ dB \cite{GuLiWuYang}) has already been achieved and as a result using a $0.66$ squeezing is of no problem. It is crucial to remember that the squeezing is generally represented in decibels and the position squeezing $s$ is related to the dimensionless squeezing parameter $r$ via the relation $s=-10\log_{10}[e^{-2r}]$\cite{Lvovsky}. In a crystal lattice, using a second order Raman scattering, phonons have been squeezed \cite{Raman1,Raman2}. For cold bosonic atoms in optical lattices \cite{Cold_Bosonic_Optical_Lattice}, such second order Raman scattering \cite{Raman1} or pump-probe detection scheme \cite{Pump_Probe_Detection} can be used to squeeze the phonon modes when an optical lattice potential is present. One can now obtain obtain the final form of eq.(\ref{1.108}) as 
\begin{widetext}
\begin{equation}\label{1.111}
\begin{split}
\frac{1}{\mathfrak{N}\langle(\Delta\varepsilon)^2\rangle}\leq\frac{V_\beta \mathfrak{r}_1}{7680\sqrt{2\pi}c_s^3\tau^3}(\Omega^4\tau^4+6\Omega^2\tau^2+3-3e^{-\frac{\Omega^2\tau^2}{2}})+\frac{\hbar GV_\beta \Omega_m^2\mathfrak{r}_1\mathcal{B}(\tau)}{225 \pi\sqrt{2\pi}\varepsilon^2c_s^3\tau^3 c^5}(\Omega^4\tau^4+6\Omega^2\tau^2+3+3e^{-\frac{\Omega^2\tau^2}{2}})
\end{split}
\end{equation}
where we have used $\mathcal{B}(\tau)\equiv \mathcal{B}(r_k,\phi_k,\tau)$. 
\end{widetext}
For the right hand side of the above equation no apprximation for the $\Omega\tau$ factor has been taken. Now, we shall investigate into the case when $\Omega\sim\Omega_m$ ($10^8$ Hz). It is important to note that the BEC in general is prepared in a single length direction and the perpendicular directions are quite smaller. Our model on the other hand carries the cubic BEC approximation. It has been possible to create a BEC with length $L_\beta\sim 10^{-3}$ m \cite{Vengalattore,Barr,Greytak}. As $\tau$ is the duration of the single measurement of the gravitational wave $\tau=\frac{v_{\text{max}}}{L_\beta}=\frac{c}{L_\beta}\sim 10^{-11}$ sec. Hence, for $\Omega\sim\Omega_m$,  $\Omega\tau\sim 10^{-3}$. As a result, $\Omega\tau\ll 1$. For a total observation time of $\tau_{\text{obs.}}$, one can run approximately $\mathfrak{N}\sim\frac{\tau_{\text{obs.}}}{\tau}$ number of observations. Under the $\Omega\tau\ll 1$ condition, eq.(\ref{1.111}) can be recast as
\begin{widetext}
\begin{equation}\label{1.112}
\begin{split}
\langle (\Delta\varepsilon)^2\rangle\gtrsim&\frac{1024
\sqrt{2\pi}c_s^3\tau^2}{\Omega^2 V_\beta\tau_{\text{obs.}}\mathfrak{r}_1}\biggr(1-\frac{1024l_p^2\Omega_m^2}{50\pi\varepsilon^2c^2}\mathcal{B}(\tau)-\frac{2048l_p^2\Omega_m^2}{75\pi\varepsilon^2c^2\Omega^2\tau^2}\mathcal{B}(\tau)\biggr)~.
\end{split}
\end{equation}
\end{widetext}
It is important to note that $(\Delta\varepsilon)^2$ can never be negative, as a result from the equality condition we can write down the minimum value for the observation time of the single measurement of the gravitational wave $\tau$ to be
\begin{equation}\label{1.113}
\begin{split}
\tau_{\text{min.}}\simeq \sqrt{\frac{2}{3\pi}}\frac{32l_p\Omega_m}{c\varepsilon\Omega}\mathcal{B}(\tau)~.
\end{split}
\end{equation}
For a vacuum state without any squeezing and $\Omega\sim\Omega_m$, $\tau$ attains its absolute minimum value, which is given by
\begin{equation}\label{1.114}
\begin{split}
\tau_{\text{min.}}^0\simeq& \sqrt{\frac{2}{3\pi}}\frac{32l_p\Omega_m}{c\varepsilon\Omega}\Bigr\rvert_{\Omega\rightarrow\Omega_m}\sim\frac{1.59\times10^{-14}}{\Omega_m}\text{ sec}\\\implies\tau_{\text{min.}}^0\simeq& 1.59\times 10^{-22}\text{ sec}.
\end{split}
\end{equation}
This is a very important result in our paper. In \cite{PhononBEC3}, it was argued that the measurement cannot be arbitrarily smaller by comparing numerical data. In our case, a complete quantum gravity calculation puts up a theoretical lower bound for the measurement time when the noise fluctuation is weighted by a Gaussian decay factor. Eq.(\ref{1.114}) reveals that the single measurement time $\tau$  must be greater than or equal to  $\tau^0_{\text{min.}}$. It is although very important to note that the Gaussian decay term in the classical part of the gravitational waves comes entirely from the template of the gravitational wave whereas in this section it is imposed by hand. Therefore, the results obtained in section (\ref{S4}) are much more plausible than this one. Although one indeed can induce such Gaussian decay mechanically into the system which shall lead to a much more complicated result than the simpler model presented here.
\subsection{BEC as a graviton detector}\label{BEC_Detector}
\begin{figure}[ht!]
\begin{center}
\includegraphics[scale=0.202]{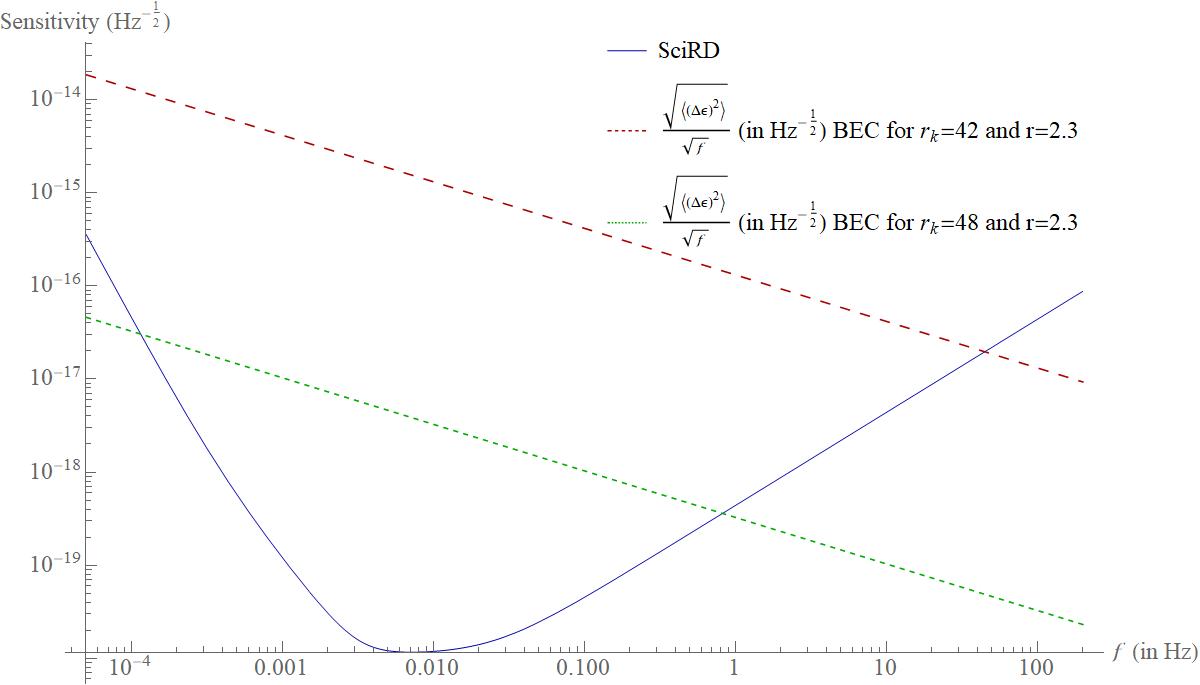}
\caption{The SciRD sensitivity formula is plotted along with  the BEC-graviton model sensitivity formula against the wave frequency $f$. The BEC sensitivity plot is viable when the resonance condition is satisfied which is $2\omega_\beta=\Omega=f$.\label{BEC_Graviton_Detector}}
\end{center}
\end{figure}

\begin{figure}[ht!]
\begin{center}
\includegraphics[scale=0.182]{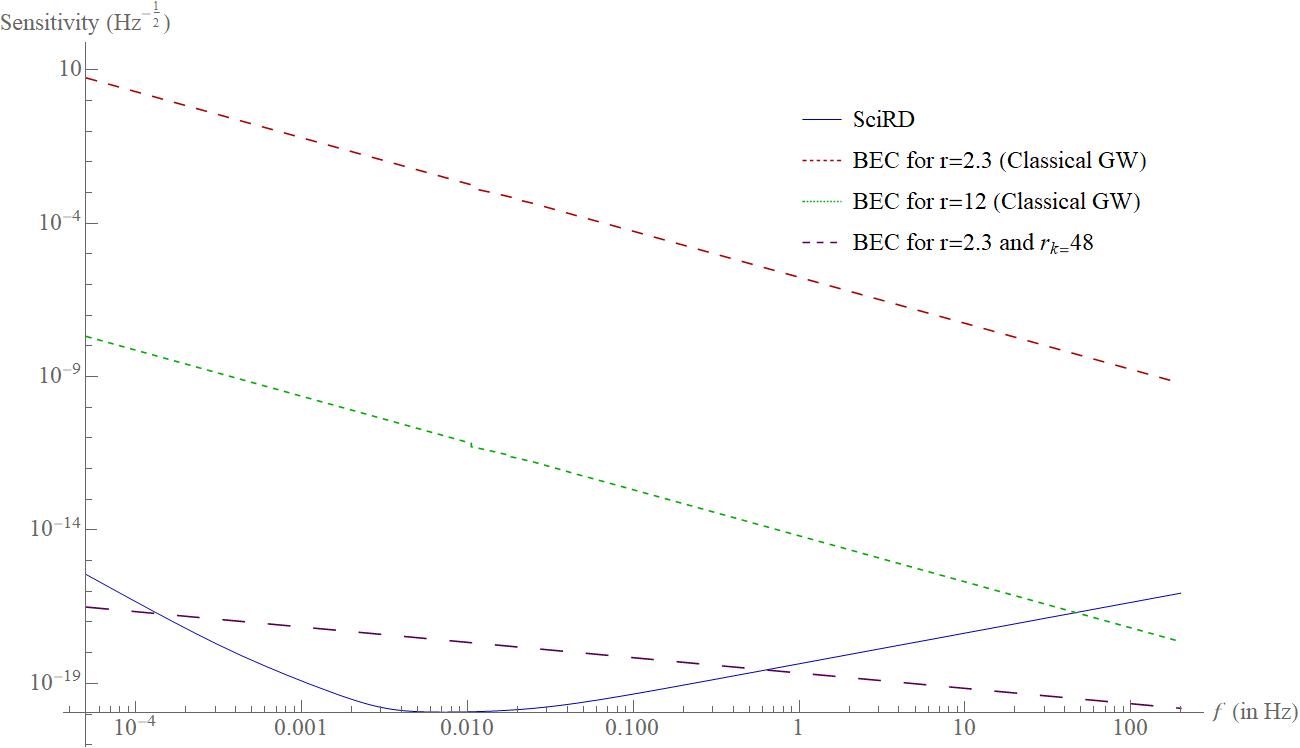}
\caption{The SciRD sensitivity formula is plotted along with  the BEC model sensitivity formula against the wave frequency $f$. We plot the case of BEC-Classical Gravity wave model alongside the BEC-Graviton model. \label{BEC_Graviton_Detector2}}
\end{center}
\end{figure}

\begin{figure}[ht!]
\begin{center}
\includegraphics[scale=0.182]{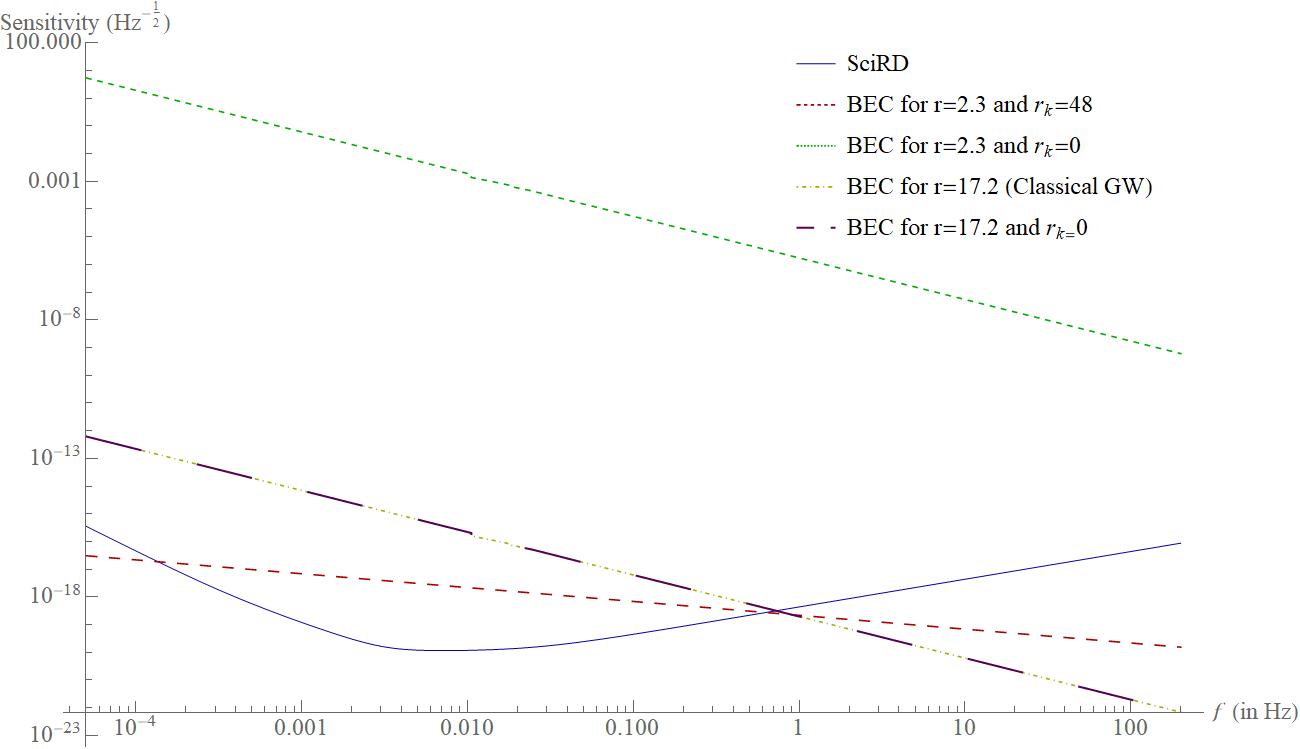}
\caption{The SciRD sensitivity formula is plotted along with  the BEC model sensitivity formula against the wave frequency $f$. We plot the case of BEC-Classical Gravity wave model alongside the BEC-Graviton model when the incoming graviton is coming with and without any initial squeezing. \label{BEC_Graviton_Detector3}}
\end{center}
\end{figure}

\noindent In this subsection, we shall argue that the BEC will suffice as a graviton detector when future generation of gravitational wave detector will come up. We shall here use the projected sensitivity of the upcoming LISA\footnote{The full form of LISA is Laser Interferometer Space Antenna.} observatory as a baseline for the comparison.
 In the next section, we shall consider a more realistic case when there are interaction between the phonon modes which will result in a decoherence effect. We start with the sensitivity formula presented in Science Requirement Document (SciRD) \cite{SciRD} projected for the LISA observatory. A detailed discussion can be obtained in \cite{SciRD2}. The SciRD sensitivity formula reads \cite{SciRD,SciRD2}
 \begin{align}
S_{\text{h,SciRD}}(f)&=\frac{10}{3}\left(\frac{S_I(f)}{(2\pi f)^4}+S_{II}(f)\right)R(f)\text{Hz}^{-1}\label{SS1}\\
S_I(f)&=5.76\times 10^{-48}\left(1+\frac{f_1^2}{f^2}\right)\text{sec}^{-4}\cdot\text{Hz}^{-1}\label{SS2}\\
S_{II}(f)&=3.6\times 10^{-41}\text{Hz}^{-1}\label{SS3}\\
R(f)&=1+\frac{f^2}{f_2^2}\label{SS4}
 \end{align}
 where $f_1=0.4\text{mHz}$ and $f_2=25\text{mHz}$.
In order to compare the above result we consider the equality from eq.(\ref{1.111}) and use the minimum value of the standard deviation in the amplitude parameter $\sqrt{\langle(\Delta\varepsilon)^2\rangle}_{\text{min}}\bigr\rvert_{\Omega=f}$. The sensitivity of the BEC is given by $\frac{\sqrt{\langle(\Delta\varepsilon)^2\rangle}_{\text{min}}}{\sqrt{f}}$ $\text{Hz}^{-\frac{1}{2}}$. We use the following parameter values $\tau=10^{-6}\text{ sec}$, $\tau_{\text{obs}}=10^2\text{ sec}$, $L_\beta=10^{-3} \text{ m}$, and $c_s=0.012 \text{ m}\cdot\text{sec}^{-1}$. The plot of $\frac{\sqrt{\langle(\Delta\varepsilon)^2\rangle}_{\text{min}}}{\sqrt{f}}$ for the BEC vs the SciRD sensitivity formula ($\sqrt{S_{\text{h,SciRD}}(f)}\text{ Hz}^{-\frac{1}{2}}$) is plotted in Fig. (\ref{BEC_Graviton_Detector}). It is important to note that LISA is mainly going to work for the detection of very low frequency gravitational waves (especially primordial gravitational waves).
From Fig.(\ref{BEC_Graviton_Detector}), it is evident that with higher squeezing from the gravitons lower frequencies can be probed by the BEC. It is important to note that SciRD plot for the LISA targets classical gravitational waves. Hence, a simultaneous detection by LISA and a BEC will prove the existence of gravitons. We now compare this SciRD sensitivity formula with the case when the BEC is interacting as a classical gravitational wave in Fig.(\ref{BEC_Graviton_Detector2}).
In Fig.(\ref{BEC_Graviton_Detector2}), we observe that the semi-classical BEC model with classical gravity wave interaction is not a good candidate for detecting low-frequency gravitational waves. On the contrary with same phonon squeezing the BEC can detect graviton signatures when the gravitons are arriving with high enough squeezing. This reinforces our result, that a BEC is one of the best candidates for capturing signature of gravitons. When a gravity wave in the common frequency range will be detected by LISA, a detection by a BEC confirms the fact that gravitons exist as a classical gravity wave will never be detected by a BEC in such low-frequency ranges with fixed  squeezing as low as $r=2.3$. It also confirms that a BEC will better serve as a graviton detector than a classical gravity wave detector. It is important to observe from Fig.(\ref{BEC_Graviton_Detector2}) that with higher squeezing of the phonons, the BEC gets more adapt towards detecting classical gravitational wave signals. In order to truly investigate the feasibility of the BEC as a graviton detector, we plot the sensitivity against the gravitational wave frequency  for the case with and without graviton squeezing along with the classical gravitational wave case in Fig.(\ref{BEC_Graviton_Detector3}).
From Fig.(\ref{BEC_Graviton_Detector3}), we observe that for no squeezing of the graviton and a phonon squeezing of $r=2.3 $, the sensitivity is very high and the BEC will be unable to perform any kind of detection. If we consider an initial graviton squeezing $r_k=48$ with phonon squeezing $r=2.3$, the BEC can detect gravitons and have been plotted in Fig.(s)(\ref{BEC_Graviton_Detector},\ref{BEC_Graviton_Detector2}). If the phonon squeezing for the BEC is very high ($r=17.2$ in Fig.(\ref{BEC_Graviton_Detector3})) then the sensitivity lies in the frequency range of $1-10$ Hz gravitational wave where the gravitons have no initial squeezing at all. But if the gravitational wave is classical in nature then also the gravitational wave is detected by such highly squeezed BEC. We can find out from Fig.(\ref{BEC_Graviton_Detector3}) that the sensitivity plot for the classical gravity wave case  as well as the graviton case with no squeezing superposes on each other. This implies the inability of the BEC to distinguish between a classical as well as a quantum gravity signal when the phonons are very highly squeezed. Hence, for the BEC to act as a perfect graviton detector, one needs to use BEC with a optimal phonon squeezing, $r\sim1-2$. It is also interesting to note that, one also does not need a very high total observation time $\tau_{\text{obs}}$ for detecting gravitons. It is therefore evident that a BEC detector, although very difficult to built, would be a nice experimental set up for detecting gravitons. This would then be the first step towards observing quantum signatures of gravity.
\section{Effects of Decoherence from interacting modes of the phonons on the ``QGFI"}\label{S6}
\noindent Upto the previous section, we have considered dissipative system only. In this section, interaction among the phonon modes will be considered as a result there will be dissipation inside of the system. The simple idea is to connect the single mode BEC system with a thermal bath. For single mode Gaussian state, the time evolution of the covariance matrix reads
\begin{equation}\label{1.115}
\begin{split}
\Sigma(t)=\Gamma(t)\Sigma_0\Gamma^T(t)+\Sigma_\infty(t)
\end{split}
\end{equation}
where $\Sigma_0$ is the covariance matrix of the single mode Gaussian state initially and $\Gamma(t)=e^{-\frac{\gamma t}{2}}\mathbb{1}_{2}$ with $\gamma$ being the dissipation constant. Here in eq.(\ref{1.115}), $\Sigma_\infty(t)$ denotes the time dependent covariance matrix of the Gaussian reservoir and is given by
\begin{equation}\label{1.116}
\Sigma_\infty(t)=(1-e^{-\gamma t})\Sigma_\infty.
\end{equation}
Using eq.(s)(\ref{1.115},\ref{1.116}), one can write down the elements of the covariance matrix as \cite{ThesisMatthew,Serafinietal}
\begin{equation}\label{1.117}
\Sigma_{ij}(t)=e^{-\gamma t}{\Sigma_0}_{ij}+(1-e^{-\gamma t}){\Sigma_\infty}_{ij}
\end{equation}
where $i,j=1,2$. In this section, we have followed the analysis presented in \cite{PhononBEC,Serafinietal}. The purity of the quantum state is given by $\mu(t)=\frac{1}{2\sqrt{\det [\Sigma(t)]}}$. In such a scenario, the elements of the covariance matrix takes the form

\begin{equation}\label{1.118}
\begin{split}
\Sigma_{11}(t)&=\frac{1}{2\mu(t)}\left(\cosh 2r(t)+\cos\varphi(t)\sinh 2r(t)\right)~,\\
\Sigma_{12}(t)&=\Sigma_{21}(t)=\frac{1}{2\mu(t)}\sin\varphi(t)\sinh 2r(t)~,\\
\Sigma_{22}(t)&=\frac{1}{2\mu(t)}\left(\cosh 2r(t)-\cos\varphi(t)\sinh 2r(t)\right)~.
\end{split}
\end{equation}
In the above equation the squeezing parameter  and squeezing phase both becomes time dependent due to dissipation in the system. One can use $\mu(0)\equiv\mu_0$ and $r(t)\equiv r_0$ to define the purity and squeezing initially of the single mode bosonic system and the elements of the initial covariance matrix read
\begin{equation}\label{1.119}
\begin{split}
{\Sigma_0}_{11}&=\frac{1}{2\mu_0}\left(\cosh 2r_0+\cos\varphi_0\sinh 2r_0\right)~,\\
{\Sigma_0}_{12}&={\Sigma_0}_{21}=\frac{1}{2\mu_0}\sin\varphi_0\sinh 2r_0~,\\
{\Sigma_0}_{22}&=\frac{1}{2\mu_0}\left(\cosh 2r_0-\cos\varphi_0\sinh 2r_0\right)
\end{split}
\end{equation}
The covariance matric elements of the Gaussian reservoir initially reads
\begin{equation}\label{1.120}
\begin{split}
{\Sigma_\infty}_{11}&=\frac{1}{2\mu_\infty}\left(\cosh 2r_\infty+\cos\varphi_\infty\sinh 2r_\infty\right)~,\\
{\Sigma_\infty}_{12}&={\Sigma_\infty}_{21}=\frac{1}{2\mu_\infty}\sin\varphi_\infty\sinh 2r_\infty~,\\
{\Sigma_\infty}_{22}&=\frac{1}{2\mu_\infty}\left(\cosh 2r_\infty-\cos\varphi_\infty\sinh 2r_\infty\right)
\end{split}
\end{equation}
where $\mu_\infty$, $r_\infty$ and $\varphi_\infty$ denote respectively the purity, squeezing parameter, and squeezing angle of the reservoir. One can easily consider a thermal bath with no squeezing which is given by the condition $r_\infty=0$ \cite{PhononBEC,ThesisMatthew} and reduces the covariance matrix of the thermal bath to $\Sigma_\infty=\frac{1}{2\mu_\infty}\mathbb{1}_2$. Initially, we shall start with non-zero squeezing for the thermal bath and later will reduce down to the no squeezing case. Using eq.(\ref{1.118},\ref{1.119},\ref{1.120}) in eq.(\ref{1.117}), one obtains three equations which are given by
\begin{widetext}
\begin{align}
\frac{\cosh 2r(t)+\cos\varphi(t)\sinh 2r(t)}{2\mu(t)}&=\frac{e^{-\gamma t}}{2\mu_0}\left(\cosh 2r_0+\cos\varphi_0\sinh 2r_0\right)+\frac{1-e^{-\gamma t}}{2\mu_\infty}\left(\cosh 2r_\infty+\cos\varphi_\infty\sinh 2r_\infty\right)\label{1.121}\\
\frac{1}{2\mu(t)}\sin\varphi(t)\sinh 2r(t)&=\frac{e^{-\gamma t}}{2\mu_0}\sin\varphi_0\sinh 2r_0+\frac{1-e^{-\gamma t}}{2\mu_\infty}\sin\varphi_\infty\sinh 2r_\infty\label{1.122}\\
\frac{\cosh 2r(t)-\cos\varphi(t)\sinh 2r(t)}{2\mu(t)}&=\frac{e^{-\gamma t}}{2\mu_0}\left(\cosh 2r_0-\cos\varphi_0\sinh 2r_0\right)+\frac{1-e^{-\gamma t}}{2\mu_\infty}\left(\cosh 2r_\infty-\cos\varphi_\infty\sinh 2r_\infty\right).\label{1.123}
\end{align}
Using the above three equations, one obtains three equations describing the dissipiation relations of the three independent parameters as \cite{Serafinietal}
\begin{align}
&\mu(t)=\mu_0\left[e^{-2\gamma t}+\frac{\mu_0^2}{\mu_\infty^2}(1-e^{-\gamma t})^2+\frac{2\mu_0}{\mu_\infty}e^{-\gamma t}(1-e^{-\gamma t})\left(\cosh 2r_0\cosh2r_\infty-\cos(\varphi_0-\varphi_\infty)\sinh 2r_0\sinh2r_\infty\right)\right]^{-\frac{1}{2}},\label{1.124}\\
&\cosh 2r(t)=\mu(t)\left[\frac{e^{-\gamma t}}{\mu_0}\cosh2r_0+\frac{1-e^{-\gamma t}}{\mu_\infty}\cosh2r_\infty\right],
~\tan\varphi(t)=\frac{\sin\varphi_0\sin2r_0+\frac{\mu_0}{\mu_\infty}(e^{\gamma t}-1)\sin\varphi_\infty\sinh 2r_\infty}{\cos\varphi_0\sin2r_0+\frac{\mu_0}{\mu_\infty}(e^{\gamma t}-1)\cos\varphi_\infty\sinh 2r_\infty}~.\label{1.125}
\end{align}
\end{widetext}
Our calculation produces results slightly different than the one presented in \cite{Serafinietal}. One of the primary reasons is that the different signature of the off-diagonal elements of the covariance matrix corresponding to the single mode Gaussian state of the Bose-Einstein condensate.
We shall now set $r_\infty=0$ which recasts the `$\varphi(t)$' equation in eq.(\ref{1.125}) to $\tan\varphi(t)=\tan\varphi_0$. This implies that the squeezing angle does not change overtime if the reservoir attached has no squeezing. Hence, we can replace $\phi(t)$ by $\phi_0$ in our analysis. Eq.(s)(\ref{1.124},\ref{1.125}), in this non-squeezed thermal bath consideration, then reduces to \cite{PhononBEC3,ThesisMatthew}
\begin{widetext}
\begin{equation}\label{1.126}
\begin{split}
\mu(t)=&\mu_0\biggr(e^{-2\gamma t}+\frac{\mu_0^2}{\mu_\infty^2}(1-e^{-\gamma t})^2+\frac{2\mu_0}{\mu_\infty}e^{-\gamma t}(1-e^{-\gamma t})
\cosh 2r_0\biggr)^{-\frac{1}{2}},~\cosh 2r(t)=\mu(t)\left[\frac{e^{-\gamma t}}{\mu_0}\cosh 2r_0+\frac{1-e^{-\gamma t}}{\mu_\infty}\right]~.
\end{split}
\end{equation}
Assuming that $r_0>\max\left[\frac{\mu_0}{\mu_\infty},\frac{\mu_\infty}{\mu_0}\right]$, one can get the value of $t$ for which the purity becomes minimum as \cite{ThesisMatthew,Serafinietal}
\begin{equation}\label{1.127}
t_{\text{min.}}=\frac{1}{\gamma}\ln\left[\frac{\mu_0^2+\mu_\infty^2-2\mu_0\mu_\infty\cosh 2r_0}{\mu_0^2-\mu_0\mu_\infty\cosh 2r_0}\right]~.
\end{equation}
Here, $t_{\text{min.}}$ serves as the characteristic decoherence time of the squeezed single-mode bosonic states. It is straightforward to understand that the way to incorporate the dissipation into the theory is to replace $r$ by $r(t)$ into the stochastic average of the QGFI from eq.(\ref{1.94}). We can rewrite the stochastic average of the QGFI from eq.(\ref{1.94}) as
\begin{equation}\label{1.128}
\begin{split}
\llangle\hat{\mathcal{H}}_\varepsilon\rrangle&=\frac{1}{32}\pi\omega_\beta^2\tau^2\left(
e^{2\omega_\beta\Omega\tau^2}-1\right)^2
e^{-\frac{\tau^2}{2}(\Omega+2\omega_\beta)^2}\left(3\cosh^22r(\tau)-2\right)+\frac{l_p^2\Omega_m^2}{15 \pi\varepsilon^2c^2}\Bigr(1-2\omega_\beta^2\tau^2+\cosh^22r(\tau)(1+6\omega_\beta^2\tau^2)\\&+6\omega_\beta\tau\cosh 2r(\tau)\sqrt{\cosh^22r(\tau)-1}\Bigr)\mathcal{B}(r_k,\phi_k,\tau)
\end{split}
\end{equation}
\end{widetext}
where we have replaced $t$ by $\tau$ in $r(t)$.
 The simplest way to incorporate dissipation into the theory is by replacing the $\cosh 2r(\tau)$ terms using eq.(\ref{1.126}). Instead of doing an analytical calculation, we need to compare the result using plots. It is important to note that Beliaev damping will be dominant at low temperature which in the zero temperature limit takes the form \cite{Beliaev,Beliaev_Original,Anderson}
\begin{equation}\label{1.129}
\gamma\simeq \frac{3}{640\pi}\frac{\hbar \omega_\beta^5}{m_\beta n_\beta c_s^5}
\end{equation}
where $n_\beta$ denotes the number density of the atoms in the BEC and $m_\beta$ denotes the mass of each individual atoms. For a Bose-Einstein condensate with a number density of $7\times 10^{20}\text{ m}^{-3}$, $c_s\simeq1.2\times10^{-2} \text{ m sec}^{-1}$\cite{Andrewsetal}. 
\begin{figure}
\begin{center}
\includegraphics[scale=0.244]{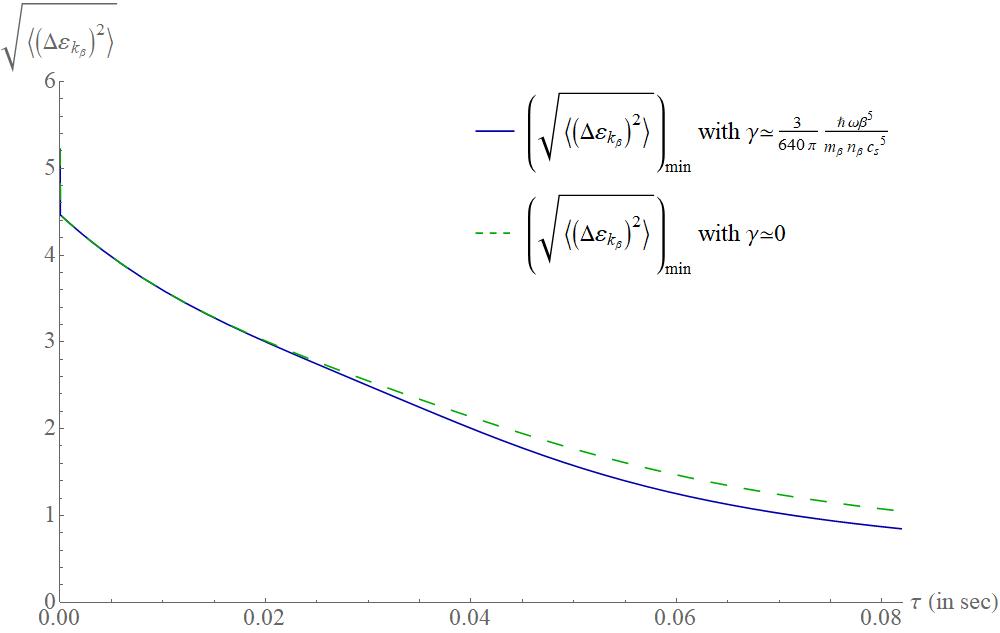}
\caption{$\sqrt{\langle(\Delta \varepsilon_{k_\beta})^2\rangle}_{\text{min.}}$ vs $\tau$ has been plotted for the case with and without damping with $r_k=33$.\label{Beliaevp1}}
\end{center}
\end{figure}

\begin{figure}[ht!]
\begin{center}
\includegraphics[scale=0.268]{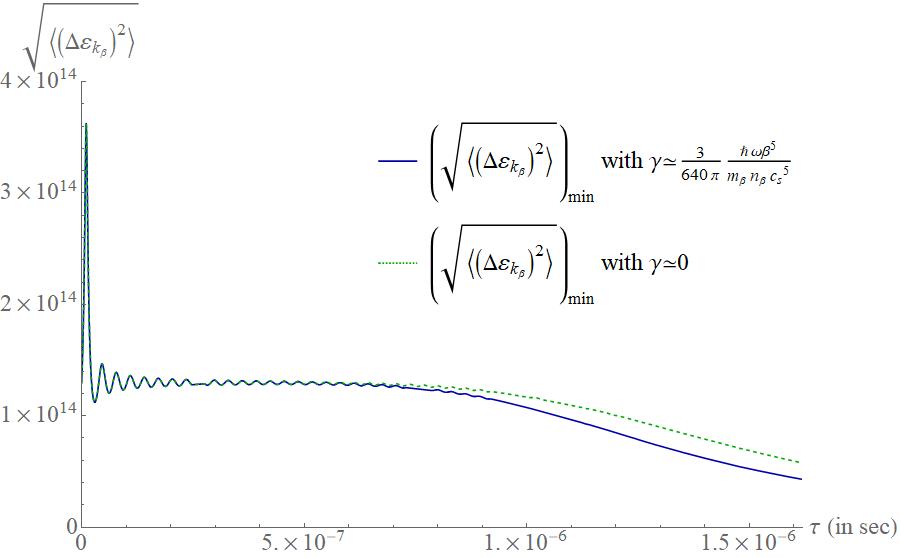}
\caption{The minimum value for the standard deviation in the amplitude parameter $\varepsilon_{k_\beta}$ has been plotted against the observation time $\tau$ for the case with and without damping with $r_k=2$.\label{Beliaevp2}}
\end{center}
\end{figure}

\noindent For $\omega_\beta= 10$ Hz, one obtains $\gamma\simeq 9.034\times 10^{-19} \text{ sec}^{-1}$. We consider the initial state of the BEC as well as the thermal bath to be pure ($\mu_0=\mu_\infty=1$). We consider mainly two cases. The first case when the squeezing of the graviton state is $r_k=33$ and the second case $r_k=2$. Both of the squeezing angles here, are set to $\frac{\pi}{2}$, the squeezing if the phonon is at $r=0.83$ and the incoming gravitational wave has a frequency $\omega=20$ Hz. For the first case we do not observe any difference due to damping untill a very later time where the decoherence results in a faster decay of the 
$\sqrt{\langle(\Delta \varepsilon_{k_\beta})^2\rangle}_{\text{min.}}$ with the observation time as can be seen from Fig.(\ref{Beliaevp1}). For Fig.(\ref{Beliaevp2}), the squeezing is reduced to $r_k=2$ and as a result $\sqrt{\langle(\Delta \varepsilon_{k_\beta})^2\rangle}_{\text{min.}}$ separates out at a very early observation time $\tau$ ($\sim 7\times 10^{-7}$ sec). This indicates that decoherence effect becomes way less significant for a lower squeezing of the initial graviton state. One can also investigate the decoherence effect for the case of the decaying noise function presented in section (\ref{S5}). For a correct incorporation of decoherence due to interacting phonon modes in the theory, one should follow the prescription in \cite{HowlFuentes}. Another important point to note is that in order to conduct such metrological measurements over a time period $\tau_{\text{obs.}}$, one needs to continuously generate the Bose-Einstein condensate using magneto-optical traps, the experimental setup of which has been proposed in \cite{Ketterle} and later observed in \cite{Tiecke}. It is important to note that we are mainly looking for signatures of quantum gravity using a Bose-Einstein condensate. In the next part of this paper we shall delve into the fundamental effects of a linearized quantum gravity theory on a BEC.

 \section{Conclusion}\label{S7}
\noindent In this paper, we have considered the simplest model of a Bose-Einstein condensate interacting with an incoming gravitational wave. The background is considered to be a flat Minkowski spacetime with fluctuations over it. In order to incorporate quantum gravitational effects into the theory, we have quantized the gravitational perturbation, over the flat background by doing a discrete Fourier mode decomposition, raising the phase space variables to operator status and applying suitable commutation relation between the conjugate variables. Using the principle of least action, an equation of motion both corresponding to the graviton as well as the time dependent part of the pseudo-Goldstone boson is obtained. Because of the involvement of the graviton part, the equation of motion corresponding to the boson becomes stochastic or Langevin-like in nature. As a result the solution obtained for the bosonic modes become stochastic as well. It is important to note that the Bogoliubov coefficients obtained in \cite{PhononBEC3} now have contributions from the noise fluctuation which raises the coefficients to an operator status. We have then used quantum metrological techniques to obtain the quantum Fisher information. Because of the quantum gravitational analysis, the quantum Fisher information picks up effects from the noise fluctuation making it stochastic in nature. This quantum gravity modified Fisher information is completely a new quantity and is termed as the ``\textit{quantum gravitational Fisher information}" (QGFI) in this paper. It is evident that the observable will be the stochastic average of the QGFI. It is important to note that we have used squeezed graviton states. The QGFI gives us very fundamental insigts into the detection scenario. From the Fig.(s)(\ref{Fish1f},\ref{Fish2f},\ref{Fish3f}) and the analytical calculation, we observe that with high enough squeezing there is a finite probabillity of the detection of a graviton background even for a very small observation time. This is not a very bizarre scenario as for a quantum gravity consideration, there is a background field always present. Hence, if a very small measurement using a single mode BEC can be done just initially, it will definitely be a graviton signature especially due to squeezed gravitons. This is the main result of our paper and it will lead completely towards a new era of graviton detection models using BEC. We have then calculated some other important aspects of such a model. We have calculated the standard deviation in the QGFI and observe that it maximizes for a higher observation time at the resonance condition. It is also possible to measure the the standard deviation of the QGFI. Although it will be very complicated, we hope for such observations in advanced experimental scenarios using continuously generated Bose-Einstein condensates. In eq.(\ref{1.94}), if we set the $l_p\rightarrow 0$ limit, we get back the result produced in \cite{PhononBEC3}. Next, we have considered a different scenario where the external noise fluctuation gets attenuated overtime by the use of a Gaussian decay factor. This analysis helps us to decay out the unusual noise fluctuations created due to the noise of gravitons. We again observe that eq.(\ref{1.112}), in the $l_p\rightarrow 0$ limit reduces to the result produced in \cite{PhononBEC3} thereby serving as a sufficient consistency check for our calculation. The Planck length dependence in our result creeps in purely due to the consideration of quantum gravity effects in our analysis. This analysis is very important in a sense that it helps us to obtain an absolute lower bound to the time of single measurement of the gravitational wave and is of the order of $10^{-22}$ sec. Next, we have used the required LISA sensitivity fromula \cite{SciRD,SciRD2} and comparing with our results, we find that a BEC will be one of the best candidates for a graviton detection. In order to detect a graviton, the graviton must come with high enough squeezing which can only exist in primordial gravitational waves coming from the inflationary time period. This is another very important observation in our paper which shows that even without high phonon squeezing \cite{PhononBEC3,PhononBEC4}, the BEC will act as a graviton detector. Finally, we have considered a more realistic scenario when the phonon modes of the Bose-Einstein condensates are interacting. We have reproduced the results of the time dependence of the purity of the Bose-Einstein condensate as well as the phonon-squeezing parameter for the covariance matrix obtained in our case. Finally, we have obtained the form of stochastic average of the QGFI when decoherence is present in the theory. In order to truly observe the behaviour of the minimum value of the standard deviation in the amplitude $\varepsilon_{k_\beta}$, we have plotted it against the single observation time for the case when decoherence is present and when decoherence is not present in the system. We have plotted for the cases of a high graviton squeezing and low graviton squeezing. It is important to note that the change in the minimum standard deviation $\varepsilon_{k_\beta}$ becomes way less significant at initial times for high enough squeezing. For almost very small (even for no squeezing case), the difference becomes significant even at initial times but the standard deviation value suggests (Fig.(\ref{Beliaevp2})) that such effects will not be observable at such initial times. This is a very complicated experimental scenario and will be very difficult to perform as the detection of the graviton signatures realizes highly on the accuracy of instantaneous measurement. Hence, the way out is to make multiple measurements and if a resonance spike is observed in the pico-nano second time regime (even micro second) from the starting of a single measurement, it shall be a conclusive evidence of the existence of a graviton. In our current analysis, we have claimed that the BEC will suffice as the best candidate as a graviton detector but for that one needs to abide by some important initial conditions. The phonon squeezing for the BEC as well as the total observation time should not be very high. From eq.(\ref{1.112}), it is evident that if the speed of sound in the BEC can be reduced then the sensitivity for the BEC increases leading to graviton detection even for gravitons with lower squeezing. For example, if the speed of sound in the BEC is $c_s=1/2\times10^{-5}$ m/sec then the BEC will detect graviton signatures in the $1$ Hz frequency range for a graviton with initial squeezing $r_k=42$. In the next part $``\textit{Zweite Abhandlung}"$, we shall explore a much more fundamental scenario where quantum gravity will play a leading role and upon experimental verification will be a conclusive evidence of the quantum nature of gravity (specifically the evidence of linearized quantum gravity)\footnote{This analysis is an extended version of the letter \cite{OTMSBEC}.}. 
\appendix
\section{Squeezed graviton state and the two point correlator}\label{AppendixA}
\noindent In this appendix, we shall calculate the two point correlator for the initial graviton state to be in a squeezed state and try to obtain eq.(\ref{1.89}) in this process. The initial graviton is considered to be in a squeezed state. If the squeezing and the displacement operators are given by 
\begin{align}
\hat{S}(r^{\text{sq.}})&=e^{\frac{1}{V}\sum\limits_{\mathbf{k},s}\left(r^{\text{sq.}*}_{k}\hat{a}_s(\mathbf{k})\hat{a}_s(-\mathbf{k})+r^{\text{sq.}}_{\mathbf{k}}\hat{a}_s^\dagger(\mathbf{k})\hat{a}_s^\dagger(-\mathbf{k})\right)},\label{A1}\\
\hat{D}(\mathfrak{B})&=e^{\frac{1}{V}\sum\limits_{\mathbf{k},s}\left(\mathfrak{B}_k\hat{a}^\dagger_s(\mathbf{k})-\mathfrak{B}_k\hat{a}_s(\mathbf{k})\right)}\label{A2}
\end{align}
where $r^{\text{sq.}}_k=r_k e^{i\phi_k}$,
then the displaced squeezed state reads
\begin{equation}\label{A3}
|r^{\text{sq.}},\mathfrak{B}\rangle=\hat{S}(r^{\text{sq.}})\hat{D}(\mathfrak{B})|0\rangle.
\end{equation}
If the Minkowski mode solution is given by $u_k(t)=\frac{1}{\sqrt{2k}}e^{-ikt}$, then the squeezed mode function has the form
\begin{equation}\label{A4}
u_k^{\text{sq.}}(t)=u_k(t)\cosh r_k-e^{-i\phi_k}u_k^*(t)\sinh r_k.
\end{equation}
Our primary aim is to calculate $\llangle \{\delta \hat{h}_I^s(\mathbf{k},t),\delta \hat{h}_I^{s'}(\mathbf{k}',t)\}\rrangle$ where the expectation is taken with respect to the state in eq.(\ref{A3}). We already know that $\hat{h}^s_I(\mathbf{k},t)=\hat{a}_s(\mathbf{k})u_k(t)+\hat{a}^\dagger_s(-\mathbf{k})u_k^*(t)$ and $\hat{\delta h}_I^s(\mathbf{k},t)=\hat{h}_I^s(\mathbf{k},t)-\langle \hat{h}_I^s(\mathbf{k},t)\rangle.$ Before proceeding further, we want to write down the following two relations (note that both the $\hat{S}$ and $\hat{D}$ operators are unitary)
\begin{align}\label{A5}
\hat{\mathcal{A}}_{\mathbf{k},s}(r^{\text{sq.}},\mathfrak{B})=\hat{D}^\dagger(\mathfrak{B})
\hat{S}^\dagger(r^{\text{sq.}})\hat{a}_s(\mathbf{k})\hat{S}(r^{\text{sq.}})\hat{D}(\mathfrak{B}).
\end{align}
It is then straightforward to obtain the following two relations by making use of eq.(s)(\ref{A1},\ref{A2}) as
\begin{widetext}
\begin{align}
\hat{\mathcal{A}}_{\mathbf{k},s}(r^{\text{sq.}},\mathfrak{B})=&(\hat{a}_s(\mathbf{k})+\mathfrak{B}_k)\cosh r_k-(\hat{a}^\dagger_s(-\mathbf{k})+\mathfrak{B}^*_k)e^{i\phi_k}\sinh r_k\label{A6}~,\\
\hat{\mathcal{A}}^\dagger_{-\mathbf{k},s}(r^{\text{sq.}},\mathfrak{B})=&(\hat{a}_s^\dagger(-\mathbf{k})+\mathfrak{B}^*_k)\cosh r_k-(\hat{a}_s(\mathbf{k})+\mathfrak{B}_k)e^{-i\phi_k}\sinh r_k\label{A7}
\end{align}
where we have made use of the fact that the sign of $k$ remains invariant for both $\mathbf{k}$ and $\mathbf{-k}$.
Using the above two relations, we obtain
\begin{equation}\label{A8}
\begin{split}
&\hat{D}^\dagger(\mathfrak{B})
\hat{S}^\dagger(r^{\text{sq.}})\hat{\delta h}_I^s(\mathbf{k},t)\hat{S}(r^{\text{sq.}})\hat{D}(\mathfrak{B})\\&=u_k(t)\hat{\mathcal{A}}_{\mathbf{k},s}(r^{\text{sq.}},\mathfrak{B})+u_k^*(t)\hat{\mathcal{A}}^\dagger_{-\mathbf{k},s}(r^{\text{sq.}},\mathfrak{B})-u_k(t)\langle\hat{\mathcal{A}}_{\mathbf{k},s}(r^{\text{sq.}},\mathfrak{B})\rangle-u_k^*(t)\langle \hat{\mathcal{A}}^\dagger_{-\mathbf{k},s}(r^{\text{sq.}},\mathfrak{B})\rangle\\&=u_k^{\text{sq.}}(t)\hat{a}_s(\mathbf{k})+u_k^{\text{sq.}*}(t)\hat{a}_s^\dagger(-\mathbf{k})
\end{split}
\end{equation}
where we have made use of eq.(\ref{A4}) to arrive at the final line of the above equation.
We already know the commutation relation among the ladder operators as $[\hat{a}_s(\mathbf{k}),\hat{a}^\dagger_s(-\mathbf{k})]=\delta_{s,s'}\delta_{\mathbf{k},-\mathbf{k}'}$. Using the above results one obtains the following relation for the two-point correlator as
\begin{equation}\label{A9}
\begin{split}
\llangle \{\delta \hat{h}_I^s(\mathbf{k},t),\delta \hat{h}_I^{s'}(\mathbf{k}',t)\}\rrangle&=\langle r^{\text{sq.}},\mathfrak{B}|\{\delta \hat{h}_I^s(\mathbf{k},t),\delta \hat{h}_I^{s'}(\mathbf{k}',t)\}|r^{\text{sq.}},\mathfrak{B}\rangle\\
&=(u_k^{\text{sq.}}(t)u_{k'}^{\text{sq.}*}(t')+u_{k'}^{\text{sq.}}(t')u_k^{\text{sq.}*}(t))\delta_{s,s'}\delta_{\mathbf{k},-\mathbf{k}'}\\
\implies\llangle \{\delta \hat{h}_I^s(\mathbf{k},t),\delta \hat{h}_I^{s'}(\mathbf{k}',t)\}\rrangle&=\delta_{s,s'}\delta_{\mathbf{k}+\mathbf{k}',0}\mathcal{Q}_{\delta h}(t,t',\mathbf{k})~.
\end{split}
\end{equation}
It is important to note from the above equation that $\mathcal{Q}_{\delta h}(t,t',\mathbf{k})=(u_k^{\text{sq.}}(t)u_{k}^{\text{sq.}*}(t')+u_{k}^{\text{sq.}}(t')u_k^{\text{sq.}*}(t))$ which can be simplified as
\begin{equation}\label{A10}
\begin{split}
\mathcal{Q}_{\delta h}(t,t',\mathbf{k})&=2\Re(u_k^{\text{sq.}}(t)u_{k'}^{\text{sq.}*}(t'))=\frac{1}{k}\left(\cos\left(k(t-t')\right)\cosh 2r_k-\cos\left(k(t+t')-\phi_k\right)\sinh 2r_k\right)
\end{split}
\end{equation}
\end{widetext}
which is eq.(\ref{1.89}) from the main text of this paper. The non-squeezing case can be reproduced from this result just by setting $r_k=0$ throughout the analysis.

\end{document}